\documentclass[pra,letterpaper,onecolumn,showpacs,superscriptaddress]{revtex4}

\usepackage{graphicx,psfrag,amsmath,amssymb,amsfonts,bbm,latexsym,color,dcolumn,bm}

\begin{document}

\title{Comprehensive rate coefficients for electron collision induced transitions in hydrogen}

\author{Daniel Vrinceanu}

\affiliation{Department of Physics, Texas Southern University, Houston, TX 77004, USA}
 
\author{Roberto Onofrio}

\affiliation{\mbox{Dipartimento di Fisica e Astronomia ``Galileo Galilei'', Universit\`a di Padova, 
Via Marzolo 8, 35131 Padova, Italy}}

\affiliation{ITAMP, Harvard-Smithsonian Center for Astrophysics, Cambridge, MA 02138, USA}

\author{Hossein R. Sadeghpour} 

\affiliation{ITAMP, Harvard-Smithsonian Center for Astrophysics, Cambridge, MA 02138, USA}

\date{\today}

\begin{abstract}
Energy-changing electron-hydrogen atom collisions are crucial to regulating the energy 
balance in astrophysical and laboratory plasmas and relevant to the formation of stellar 
atmospheres, recombination in H-II clouds, primordial recombination, three-body recombination 
and heating in ultracold and fusion plasmas. Computational modeling of electron-hydrogen 
collision has been attempted through quantum mechanical scattering state-to-state calculations 
of transitions involving low-lying energy levels in hydrogen (with principal quantum number $n < 7$) 
and at large principal quantum numbers using classical trajectory techniques.
Analytical expressions are proposed which interpolates the current quantum mechanical and 
classical trajectory results for electron-hydrogen scattering in the entire range of energy 
levels, for nearly all temperature range of interest in astrophysical environments.
An asymptotic expression for the Born cross-section is interpolated with a modified expression 
derived previously for electron-hydrogen scattering in the Rydberg regime using classical 
trajectory Monte Carlo simulations.
The derived formula is compared to existing numerical data for transitions involving 
low principal quantum numbers, and the dependence of the deviations upon temperature is discussed.
\end{abstract}

\keywords{atomic processes  --- stellar astrophysics --- photosphere --- early universe}

\maketitle

\section{Introduction}

Energy-changing electron-hydrogen atom collisions are relevant in several areas, among these the   
formation of stellar atmospheres \citep{mashonkina}, radio emission in 
recombination processes of H-II clouds, primordial cosmological recombination of 
hydrogen \citep{chluba}, and plasma fusion physics \citep{janev}. 
Although elastic scattering and excitation of low-lying atomic states in collisions of electrons 
with ground state hydrogen atoms have been extensively studied theoretically and experimentally 
\citep{janev}, the literature for transitions among high-lying Rydberg states is scarce. 
Little direct experimental data is available for e$^-$-H(n) collisional excitation and 
transition into Rydberg states \citep{rolfes,nagesha}. 
Therefore, for these processes one has to rely mainly on theory and, due to the 
various approximation schemes, accuracy among the results within a factor of 2 
is considered excellent \citep{przybilla}.

For electron-impact transitions of low Rydberg states ($n < 7$) 
quantum mechanical methods, such as the R-matrix method \citep{anderson} 
and convergent close coupling (CCC) \citep{bray}, can be relied upon to produce 
reasonably accurate cross sections and rate coefficients. 
At the other end of the spectrum, for highly excitated states near the ionization threshold, 
the Bohr correspondence principle allows for statistical classical trajectory techniques 
\citep{mansbach} to obtain precise rate constants for such collisions. 
However, in the intermediate $n$ range, satisfactory models do not exist. 
Quantum mechanical calculations become exceedingly difficult, as the size of basis 
sets grows exponentially, and classical techniques obviously fail because they do not 
take into account quantum effects. 

It is the aim of this work to provide expressions for the collision-induced rate coefficients 
which properly bridge this intermediate gap. To this end, we employ expressions for the rate 
coefficients from Pohl, Vrinceanu \& Sadeghpour \citep{pohl}, developed for small energy-transfer 
(small $\Delta n$ transitions) low-temperature processes in ultracold laboratory plasmas, and 
analytically extend these expressions for applications to high temperature conditions and in 
agreement with quantum mechanical data for e$^-$-H(n) energy-changing collisions. 
The new recommended expressions are globally accurate for electron impact transitions 
in hydrogen atoms, and for both small-energy and large-energy transfers. 

Theoretical models based on classical mechanics have the advantage that the resulting collision 
cross sections and rates obey simple scaling laws. Therefore, based on the correspondence 
principle, their validity is expected to hold for states with large principal quantum number, 
described by classical orbits that satisfy Bohr's quantization condition. 
Early simple Thomson models with a frozen target were followed by applications of classical 
perturbation theory \citep{gryzinski}, impulse approximation \citep{gerjuoy}, 
and binary encounter \citep{flannery}. Statistical methods follow ensemble of trajectories 
to provide exact classical results. Classical Trajectory Monte Carlo (CTMC) simulations were 
introduced \citep{percival}, and applied to transition state theory \citep{mansbach}, and 
extensive results were obtained in \citet{vrinceanu}, and \citet{pohl}.
However, the classical approach is not expected to give reliable results at low energies, 
close to the  reaction threshold for excitation (although surprisingly good results can 
be obtained, as demonstrated in \citet{wannier}), and it fails to reproduce the 
characteristic high energy $\log(E)/E$ behavior of the cross section, predicting instead a $1/E$ 
behaviour \citep{beigman}.

A rigorous quantum mechanical approach must include the Born approximation as a limiting case 
\citep{bethe}. Expressions can be derived for transitions between any Rydberg states by 
involving the atom inelastic form factor, which describe satisfactorily the {\em impulsive} 
part of the interaction of the electron projectile with the Rydberg atom.
For the case of dipole (optically allowed) transitions, it correctly 
predicts the logarithmic asymptotic cross section at large energy.
A Born approximation in the impact parameter representation \cite{seaton} 
describes accurately the {\em long range} part of the interaction, but has issues  
in preserving unitarity. Large scale R-matrix \citep{anderson}, 
Convergent Close Coupling (CCC) \citep{bray}, and time dependent 
coupled channel \citep{pindzola} methods provide the most 
precise results. However, for increasing quantum numbers the difficulty of calculation 
grows exponentially. Reliable results have been obtained only for 
$ n \lesssim 5-6$ due to the complexity in accurately representing the target state 
and the number of partial waves (channels) required for convergence of results. 
It is also possible to treat the collision from a purely radiative point of view, where 
the projectile electron interacts with the Rydberg atom by exchanging equivalent (virtual) 
photons \citep{weizsacker,williams}. 
Semi-empirical and fit formulae were proposed, see for example \citep{omidvar,gee,johnson,beigman,janev}, 
trying to provide formulas that are at the same time simple enough and relatively accurate 
when compared with the best available approximate results. 

\section{Electron-Rydberg atom collisions}

The rate at which electrons collide and induce transitions from the  
state $i$ to the state $f$ of a Rydberg atom is calculated according 
to the formula
\begin{equation}
R_{i\rightarrow f} = N_e \langle v\rangle \int_{\Delta E/kT}^\infty x e^{-x} Q_{i\rightarrow f}(x)\;dx,
\label{rateD}
\end{equation}
where $N_e$ is the number density of electron gas with temperature $T$ and average 
velocity $\langle v \rangle = \sqrt{8k_B T/\pi m}$, $\Delta E$ is the difference 
in energy between the states $j$ and $f$, $x=E/(k_BT)$ is the scaled energy of 
the electron projectile, and $Q_{i\rightarrow f}$ the scattering cross section.   
The last two factors in Eq. (\ref{rateD}) constitute the rate coefficient 
$k_{if}$ such that the rate is written simply as $R_{i\rightarrow f} = N_e k_{if}$. 
In the case of excitation, the cross section $Q_{i\rightarrow f}$ is finite at 
threshold because of dipole coupling between the degenerate states. 
For a high energy projectile, the cross section is dominated by a logarithmic term 
which has slowest decrease in the following asymptotic expansion \citep{bethe} (for $E > \Delta E$)
\begin{equation}
Q_{i\rightarrow f}(E) = \frac{A}{E} \log\frac{E}{\Delta E} + \frac{B}{E} + \frac{C}{E^2} + \cdots
\label{Born}
\end{equation}
The origin of the slow logarithmic term is the optically allowed transition 
between states $i$ and $f$. When no optical transition is allowed between 
states $i$ and $f$, then the cross section has the inverse energy $1/E$ decrease.
Asymptotically, at large energies, the Born cross section Eq. (\ref{Born}) is exact 
and coefficients $A$, $B$ and $C$ depend on the initial and final states. 

Substituting the Born cross section Eq. (\ref{Born}) into Eq. (\ref{rateD}), one obtains the 
Born rate coefficient that is dominated by the ``quantum factor" $\Gamma(0,z)$, 
where $z=\Delta E/kT$, which may be expressed in terms of the incomplete gamma function 
\begin{equation}
\Gamma(0,z) = \int_z^\infty \frac{e^{-x}}{x}\; dx.
\label{qf}
\end{equation}
There are several other forms encountered in literature for this integral, relating to
the exponential integral functions $E_n(z)$ and $Ei(z)$. For small arguments, {\it i.e.}
large temperatures, $T\rightarrow \infty$, the incomplete gamma has the expansion
\begin{equation}
\Gamma(0,z) = -\gamma - \log(z) + z + {\cal O}(z^2),
\label{s1}
\end{equation}
where $\gamma$ is Euler's constant. 
For large arguments $z$, the incomplete gamma can be instead approximated as
\begin{equation}
\Gamma(0,z) = \frac{e^{-z}}{z} + {\cal O}(1/z^2).
\label{s2}
\end{equation}

CTMC calculations \citep{pohl} demonstrated that 
while previous rate coefficients obtained by Mansbach and Keck \citep{mansbach} 
are correct for large energy transfers, significant corrections, singular 
in $1/\Delta E$, have to be introduced for the proper description of 
collisions at small energy transfer. 
For excitation collisions, the proposed formula is 
\begin{equation}
k_{if} = k_0 \epsilon_f^{3/2} \left[
\frac{22}{(\epsilon_i + 0.9)^{7/3}} + \frac{9/2}{\epsilon_i^{5/2} \Delta\epsilon^{4/3}}
\right] e^{\epsilon_f - \epsilon_i},
\label{P1}
\end{equation}
where $k_0 =e^4/(k_B T \sqrt{m {\cal R}})$, and $\epsilon_{i} = {\cal R}/({n_{i}}^2 k_B T)$, 
$\epsilon_{f} = {\cal R}/({n_{f}}^2 k_B T)$, with ${\cal R}$ the Rydberg constant, and 
$\Delta \epsilon=(E_f-E_i)/(k_B T)$.

It is clear that Eq. (\ref{P1}) does not describe correctly the collision rates in the limit
of large temperature, because it has a power-like $T^{-s}$ decay as opposed to
the much slower decrease $\log(T)$ suggested by Eq.~\ref{s1}. The Born inspired 
``quantum factor" (\ref{qf}) has to be incorporated into the classical formula (\ref{P1})
in order to obtain a better representation for the collision rates over the whole range
of temperatures. A simple way to accomplish this is to use Eq. (\ref{s2}) to obtain the
{\emph{classical limit}} of the ``quantum factor" (\ref{qf}) as
$$(\epsilon_i - \epsilon_f) \;\Gamma(0,\epsilon_i - \epsilon_f) \rightarrow 
\exp(\epsilon_f - \epsilon_i).$$
This suggests that by replacing the exponential factor $\exp(\epsilon_f - \epsilon_i)$ in formula
(\ref{P1}) with the ``quantum factor" $\Delta\epsilon\; \Gamma(0,\Delta\epsilon)$, one
obtains a formula that has the correct behavior at both low and large temperatures, and large
quantum numbers $n$. The last stage is now to extend the corrected formula to 
low quantum numbers. To this end, a simple fitting factor, that is in the range 
of unity uniformly across all the parameters, can be found by direct comparison 
with the accurate R-matrix results obtained by Pryzbilla and Butler for transitions between 
low quantum numbers \citep{przybilla}.

The resulting {\emph{comprehensive expression}} can now be applied 
for low and large quantum numbers and over a wide range of temperatures
\begin{equation}
k_{if} =  k_0 \left(\frac{\epsilon_f}{\epsilon_i}\right)^{3/2}  
\left[\frac{22}{(\epsilon_i + 0.9)^{7/3}} + \frac{9/2}{\epsilon_i^{5/2} \Delta\epsilon^{4/3}}\right] 
\left(\frac{3.5 + 0.18 n_f^2}{1 + 1/\epsilon_i^{5/2}}\right)\Delta\epsilon \;\Gamma(0,\Delta\epsilon).
\label{P2}
\end{equation}
The plots in Figure 1, and Tables 1 and 2, show the comparison of the proposed formula 
with the R-matrix calculation, allowing for a direct comparison with the tables present in 
\citet{przybilla}. 
In the same graphs of Figure 1 are also shown Johnson's fitting formula \citep{johnson}, 
together with the semi-empirical rates recommended by Beigman and Lebedev \citep{beigman}, 
and the original rates by Mansbach and Keck rates \citep{mansbach}.
Both plots and tables show a progressive agreement with the R-matrix 
calculation of Pryzbilla and Butler \citep{przybilla} 
at increasing $\Delta n$, already visible for $\Delta n=2$. 
In the Rydberg limit of large $n_i, n_f$, the correction term to Eq. (\ref{P1}), 
$(3.5+0.18 n_f^2)/[\epsilon_i^{3/2}(1+1/\epsilon_i^{5/2})]$ in Eq. (\ref{P2}), 
becomes $\simeq 0.18 (n_f/n_i)^2 {\cal R}/(k_B T)$ which, for lowest energy Rydberg transitions 
with $n_f \simeq n_i$, becomes of order unity at temperatures $T \simeq 0.18 {\cal R}/k_B \simeq 
3 \times 10^4$ K, {\it i.e.} in the middle of the temperature range of interest in our considerations.

\section{Conclusions}
We have introduced a formula for electron-hydrogen collision rates which interpolates between 
{\it ab-initio} evaluations at the lowest energies and Monte Carlo simulations in the 
Rydberg regime, confronting the outcome with formulas already available in the literature 
and discussing the range of temperatures in which we can assume its validity. 
Among possible applications, stellar astrophysics and the precision study of cosmic 
background radiation in the Planck era are two of the areas which can most benefit from 
this discussion. The use of {\it ab-initio} data is of great relevance in stellar astrophysics because, 
unlike former attempts \citep{mihalas}, it is then possible to reproduce 
stellar spectra in both the optical and infrared range for a broad variety of temperatures.
Also, in observational cosmology the accuracy in the knowledge of collisional rates limits 
the precision in predicting the recombination spectrum at frequencies smaller than 1 GHz 
\citep{chluba}, especially due to the rearrangement of the populations in each energy 
manifold due to angular momentum changing collisions \citep{pengelly,vrinceanu1}. 

\acknowledgments
DV is grateful to Texas Southern University High Performance Computing Center for making the necessary 
computational resources available, and to the National Science Foundation for the support received 
through a grant for the Center for Research on Complex Network at Texas Southern University (HRD-1137732). 
This work was also partially supported by the National Science Foundation through a grant for the Institute 
for Theoretical Atomic, Molecular and Optical Physics at Harvard University, and the Smithsonian 
Astrophysical Laboratory.


\clearpage

\newcommand{\figwidth}{1.50in}
\begin{figure*}
\centering
\includegraphics[width=\figwidth]{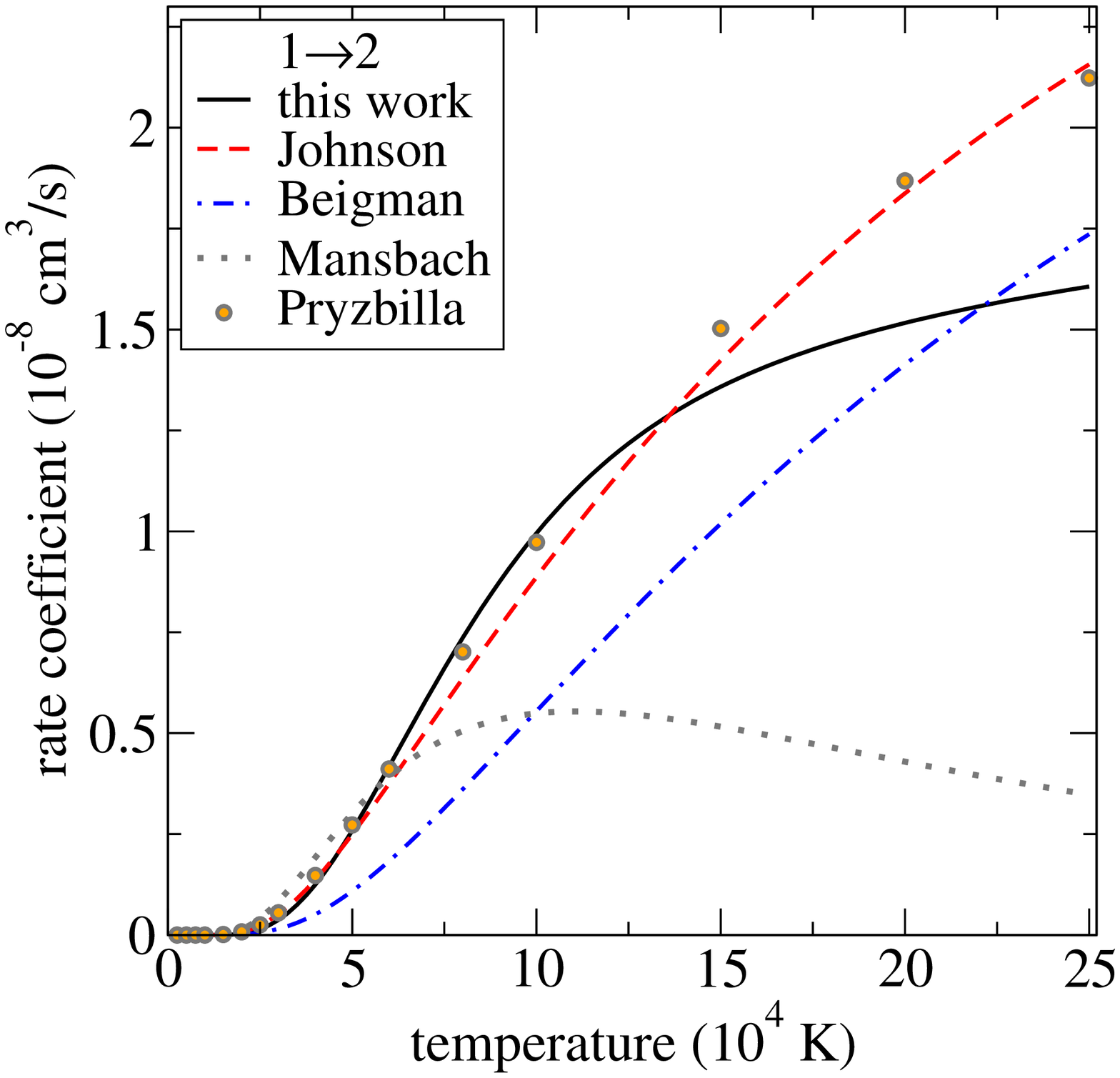}
\includegraphics[width=\figwidth]{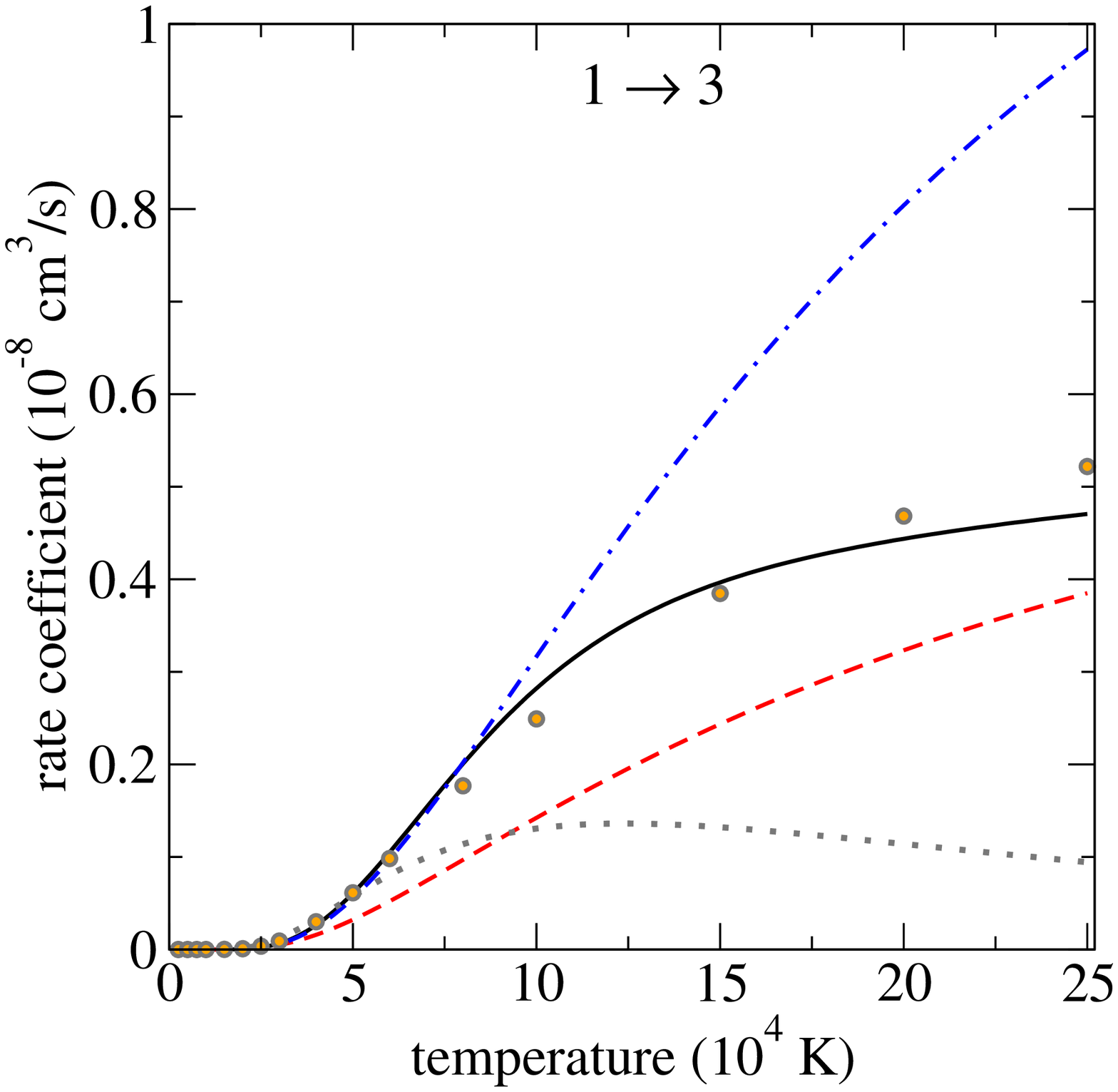}
\includegraphics[width=\figwidth]{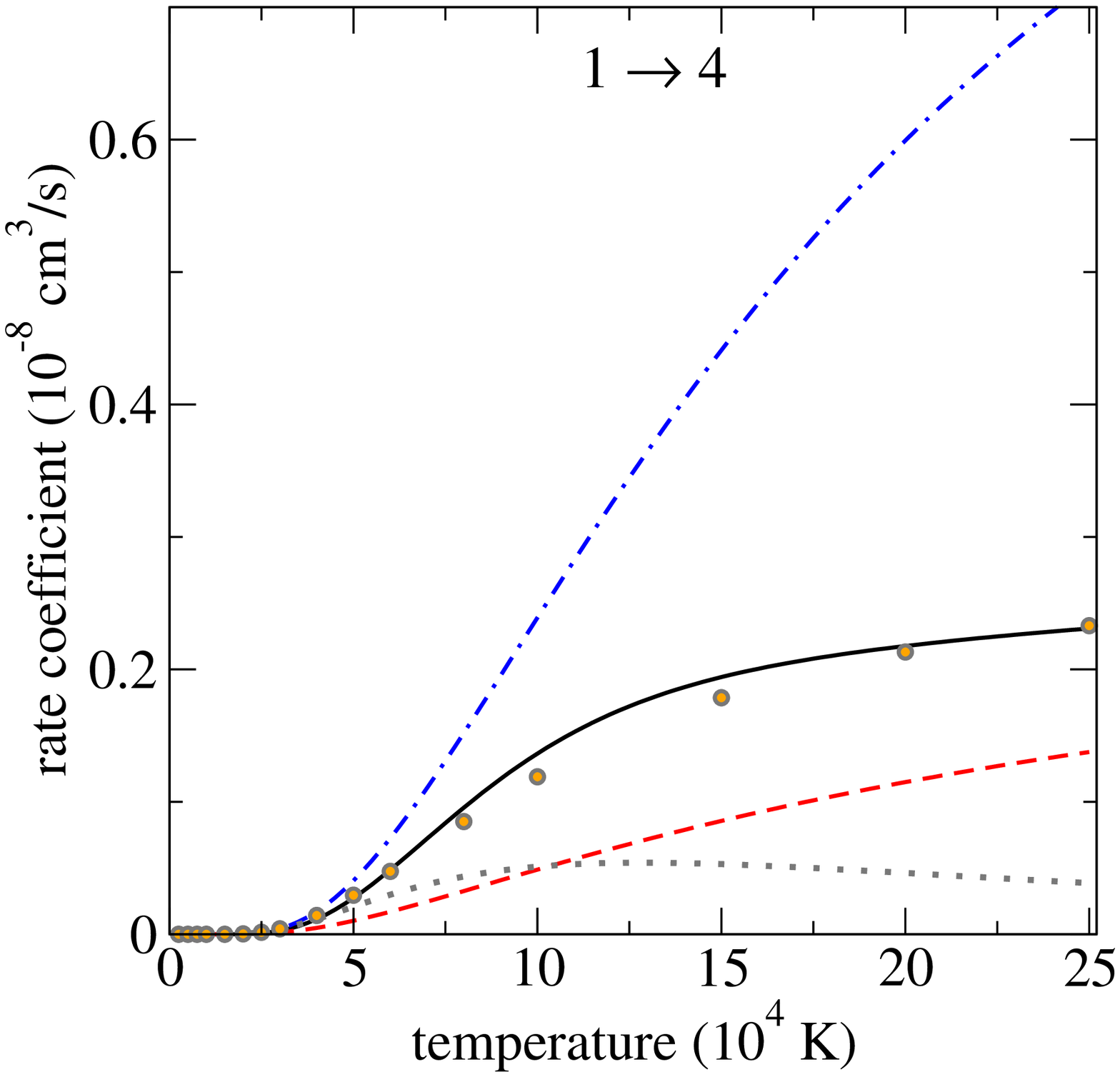}
\includegraphics[width=\figwidth]{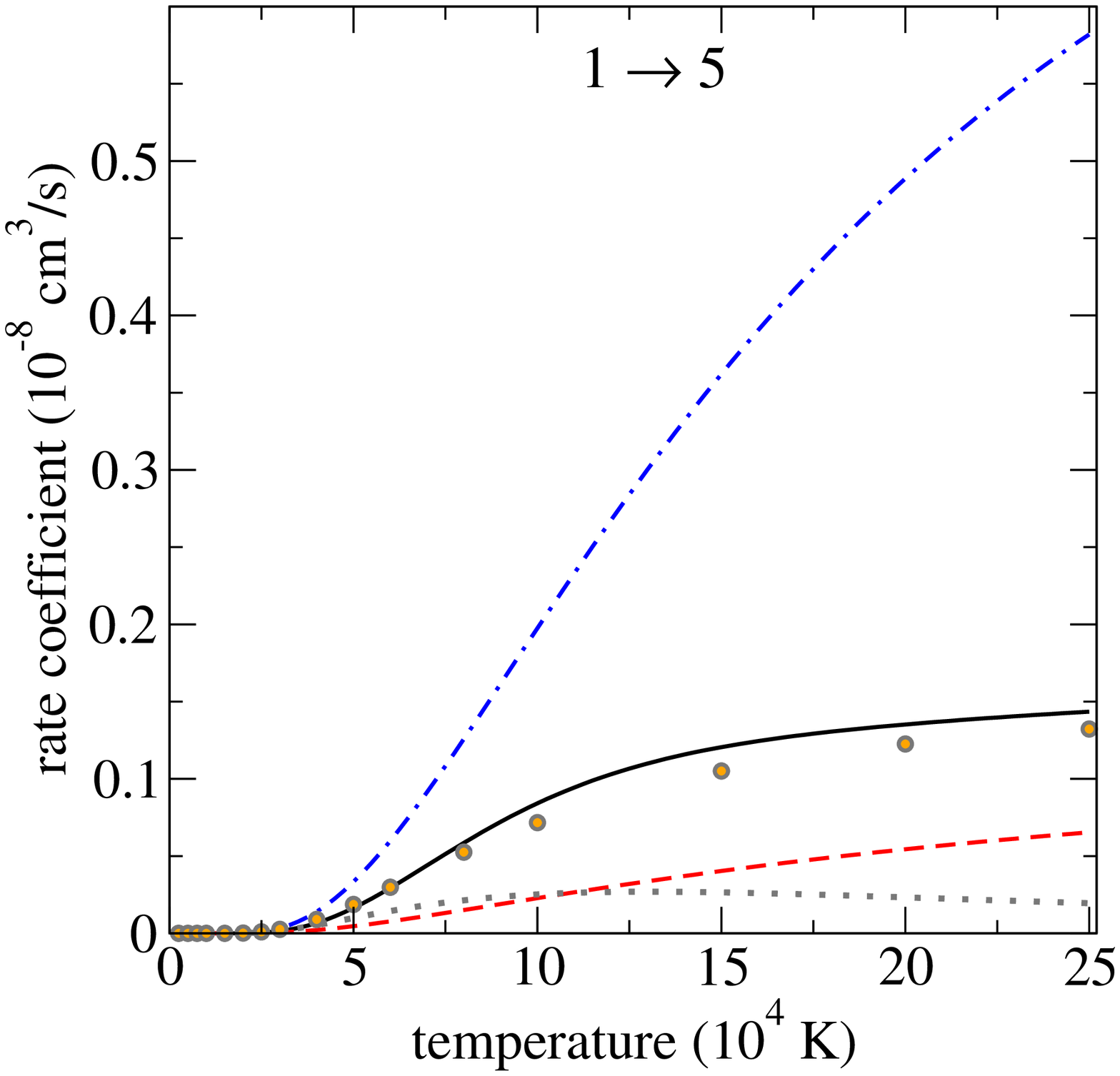}
\includegraphics[width=\figwidth]{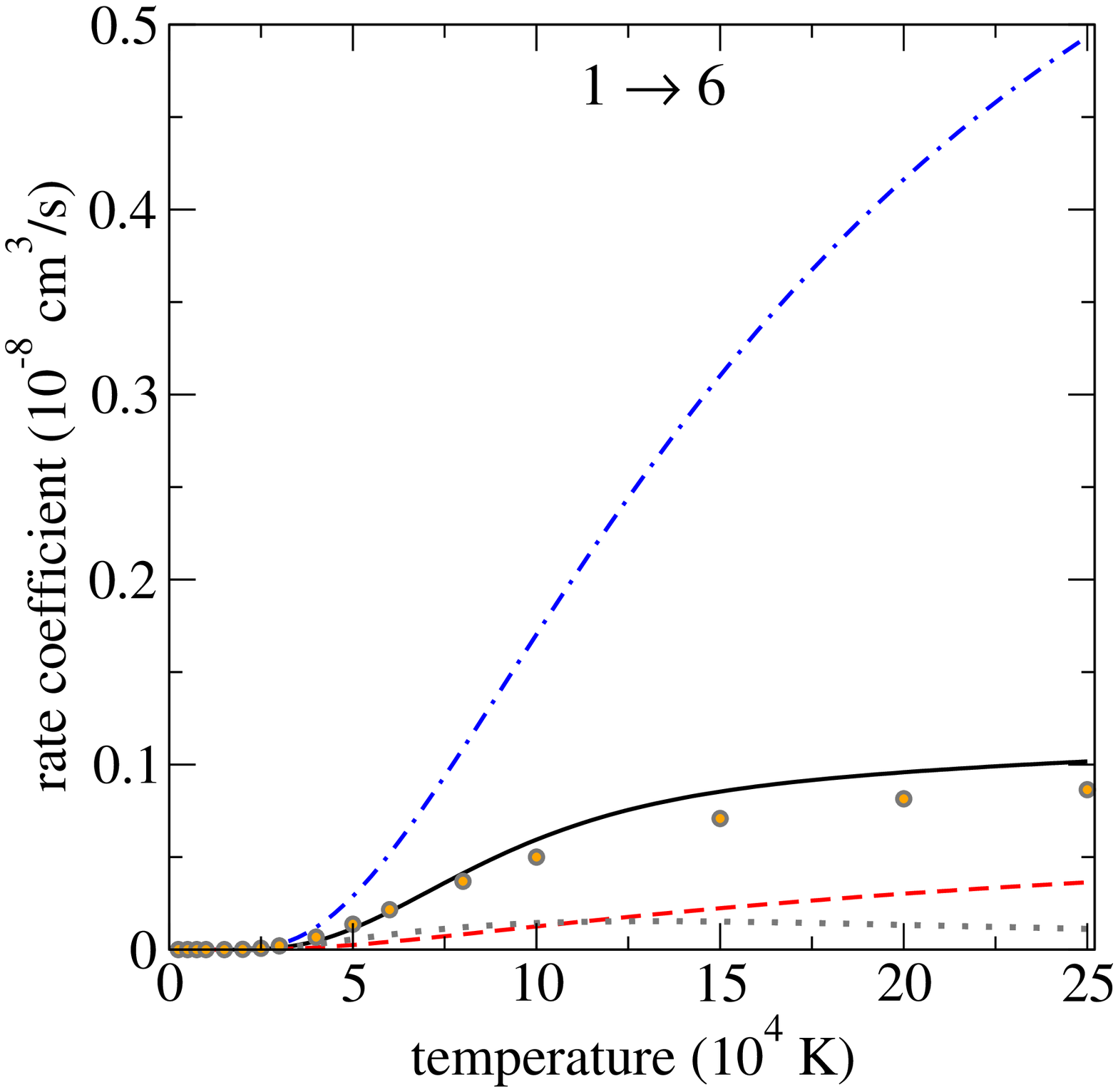}
\includegraphics[width=\figwidth]{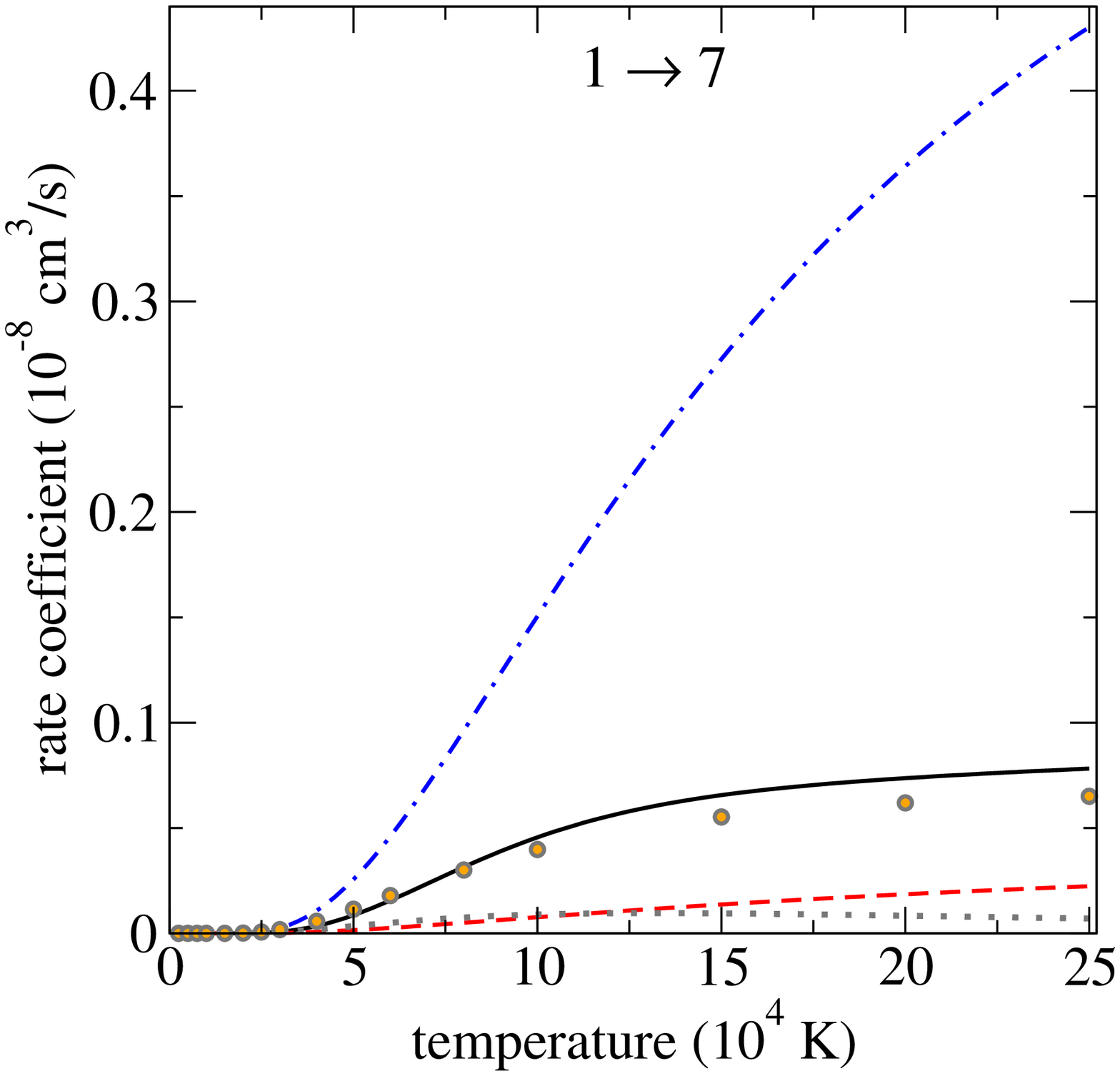}
\includegraphics[width=\figwidth]{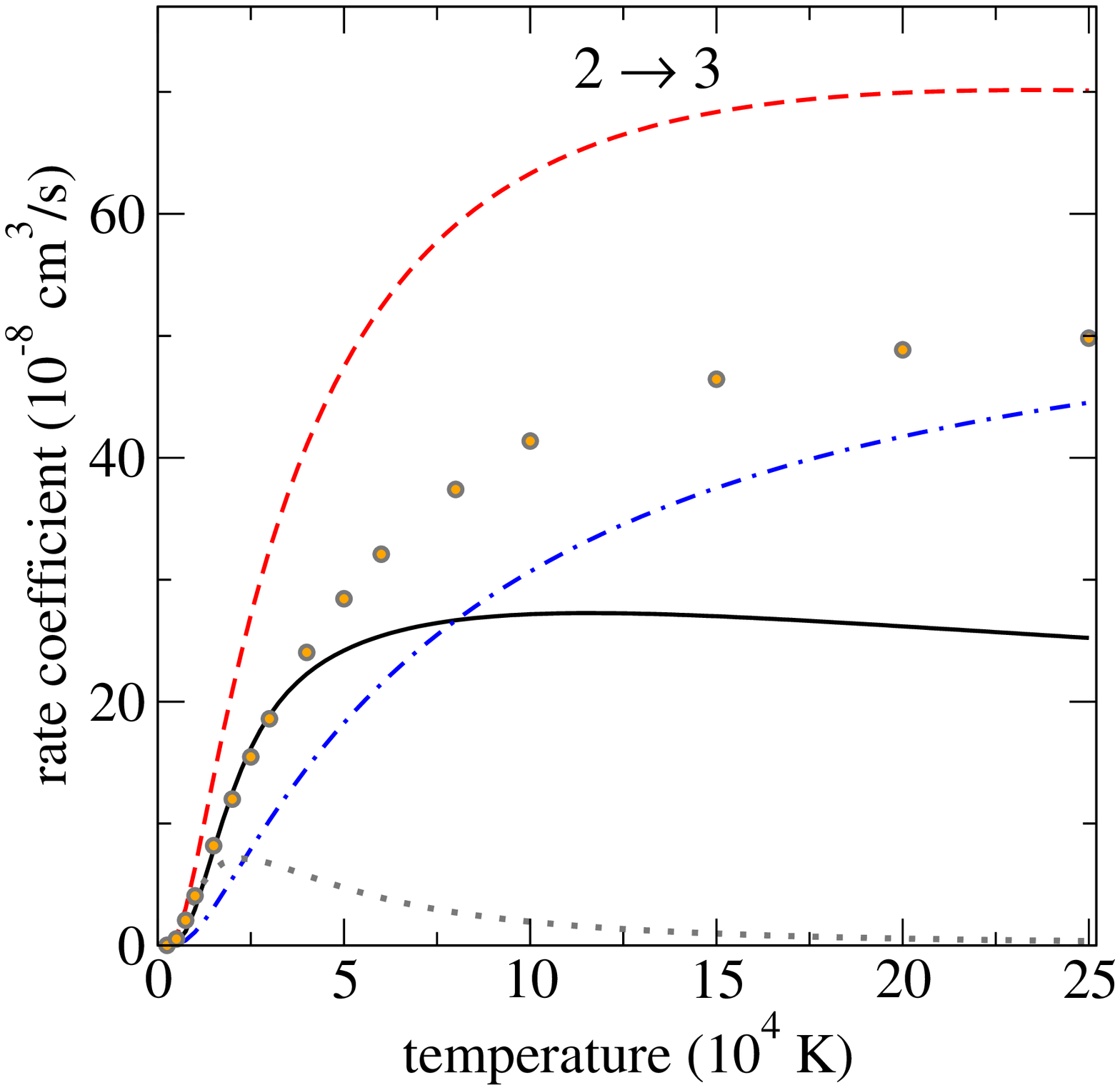}
\includegraphics[width=\figwidth]{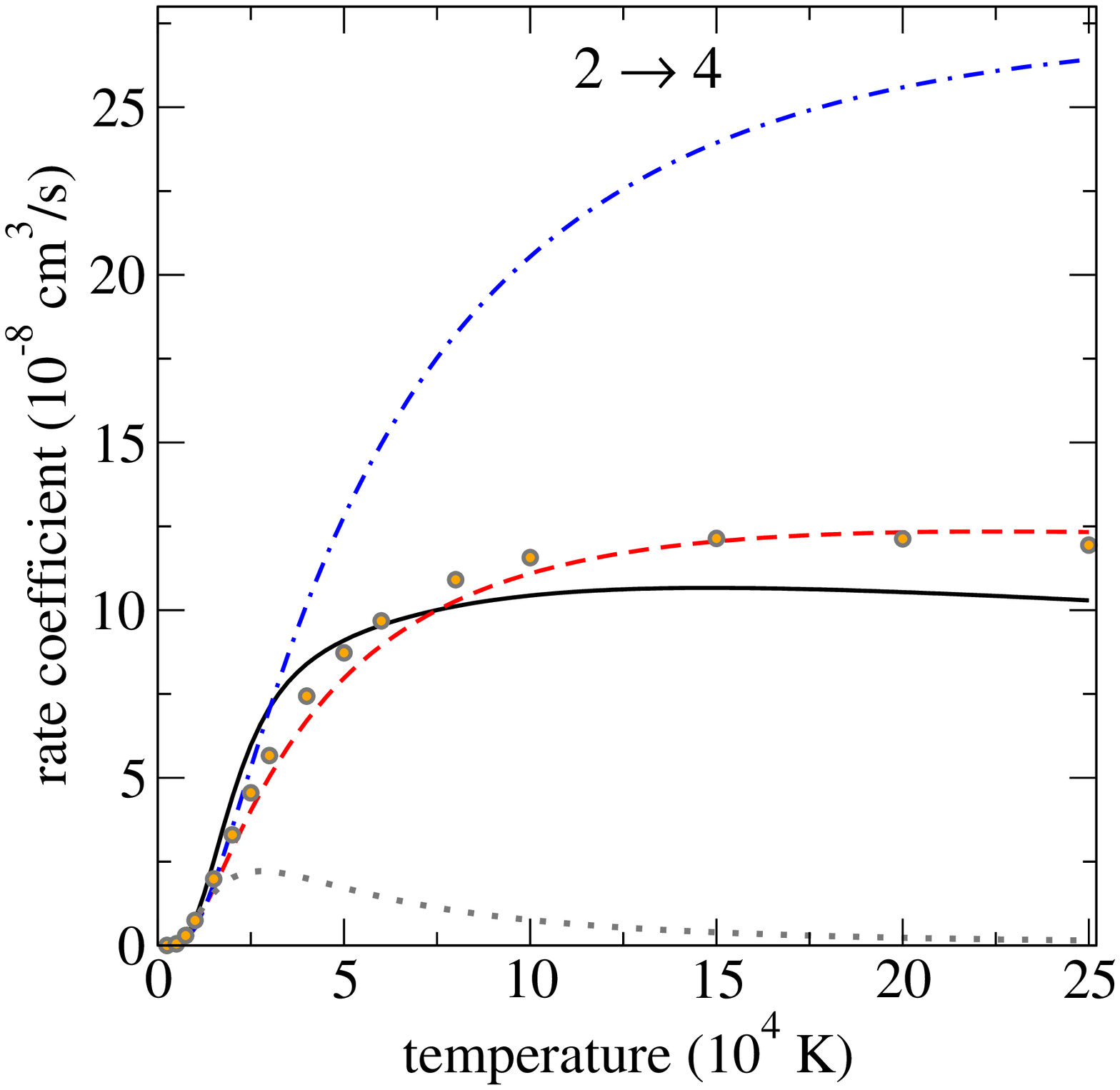}
\includegraphics[width=\figwidth]{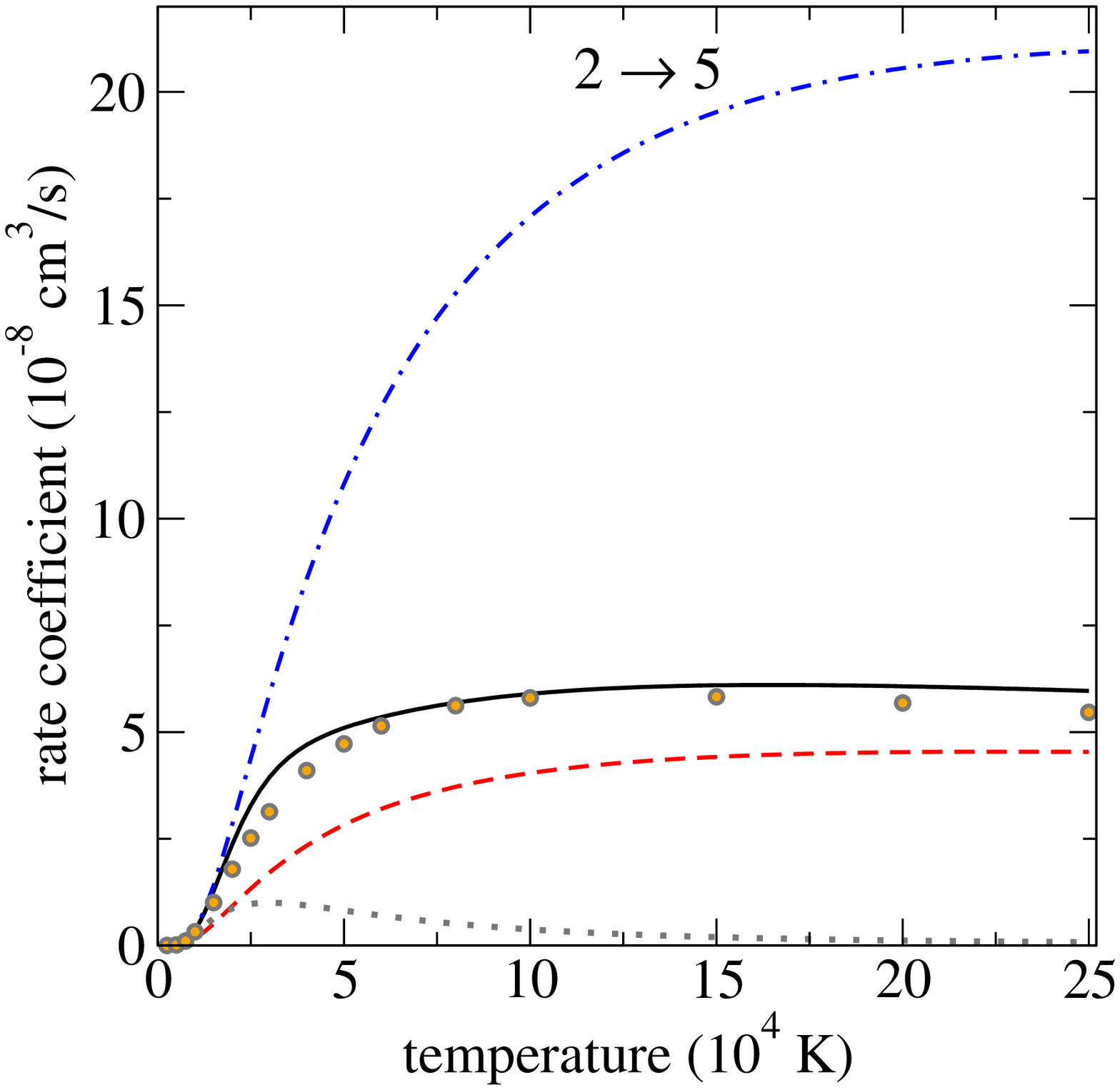}
\includegraphics[width=\figwidth]{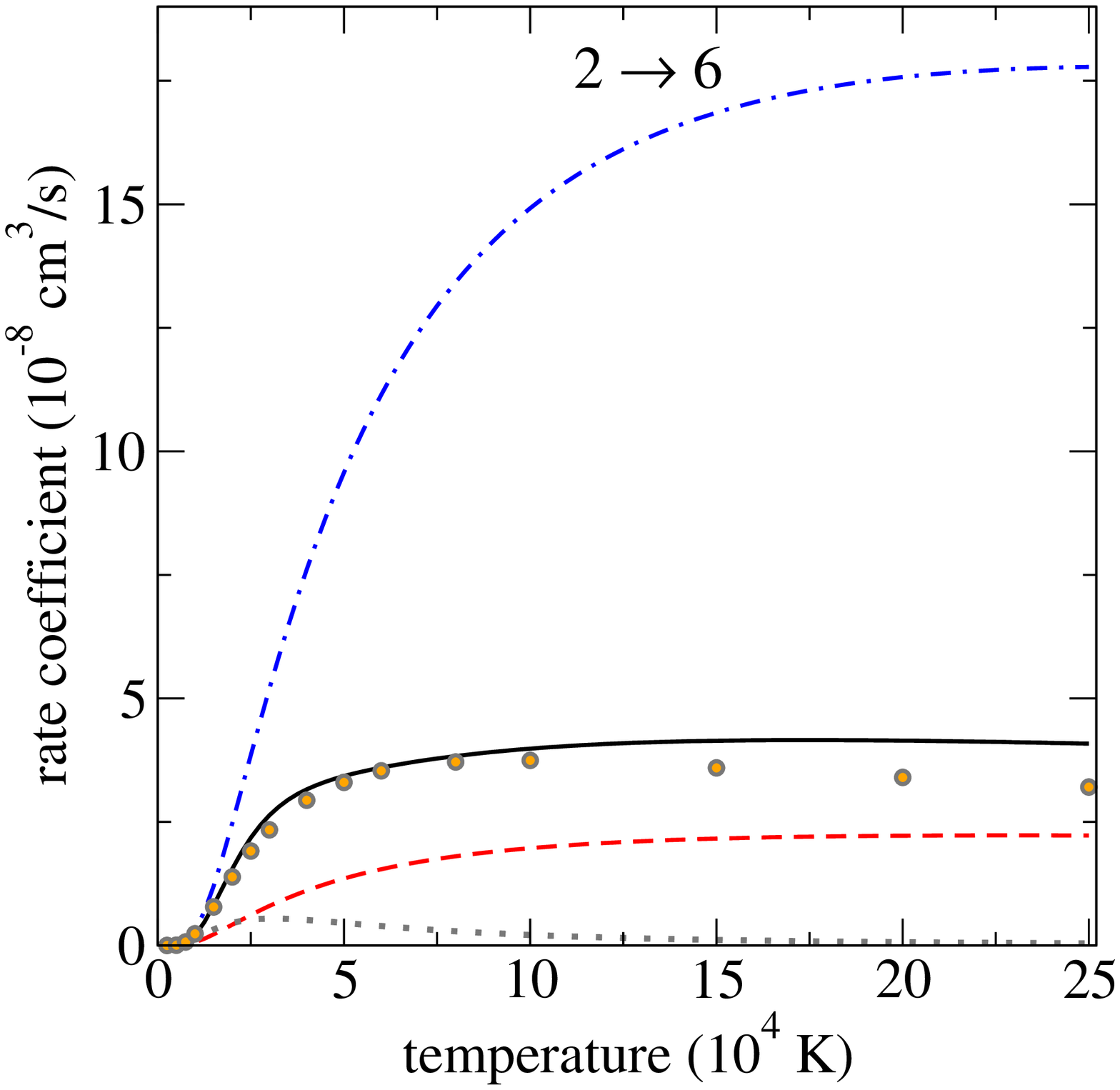}
\includegraphics[width=\figwidth]{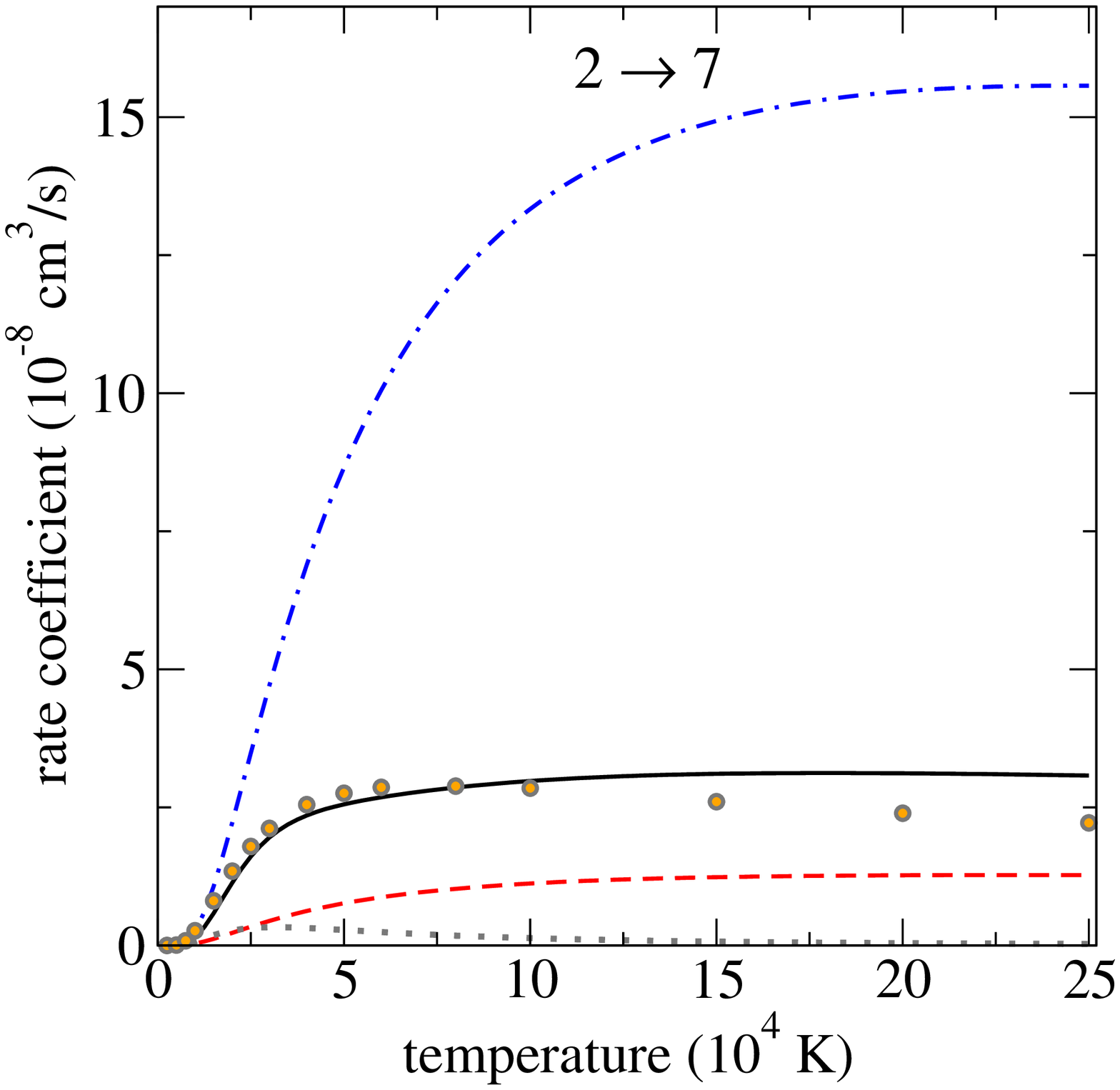}
\includegraphics[width=\figwidth]{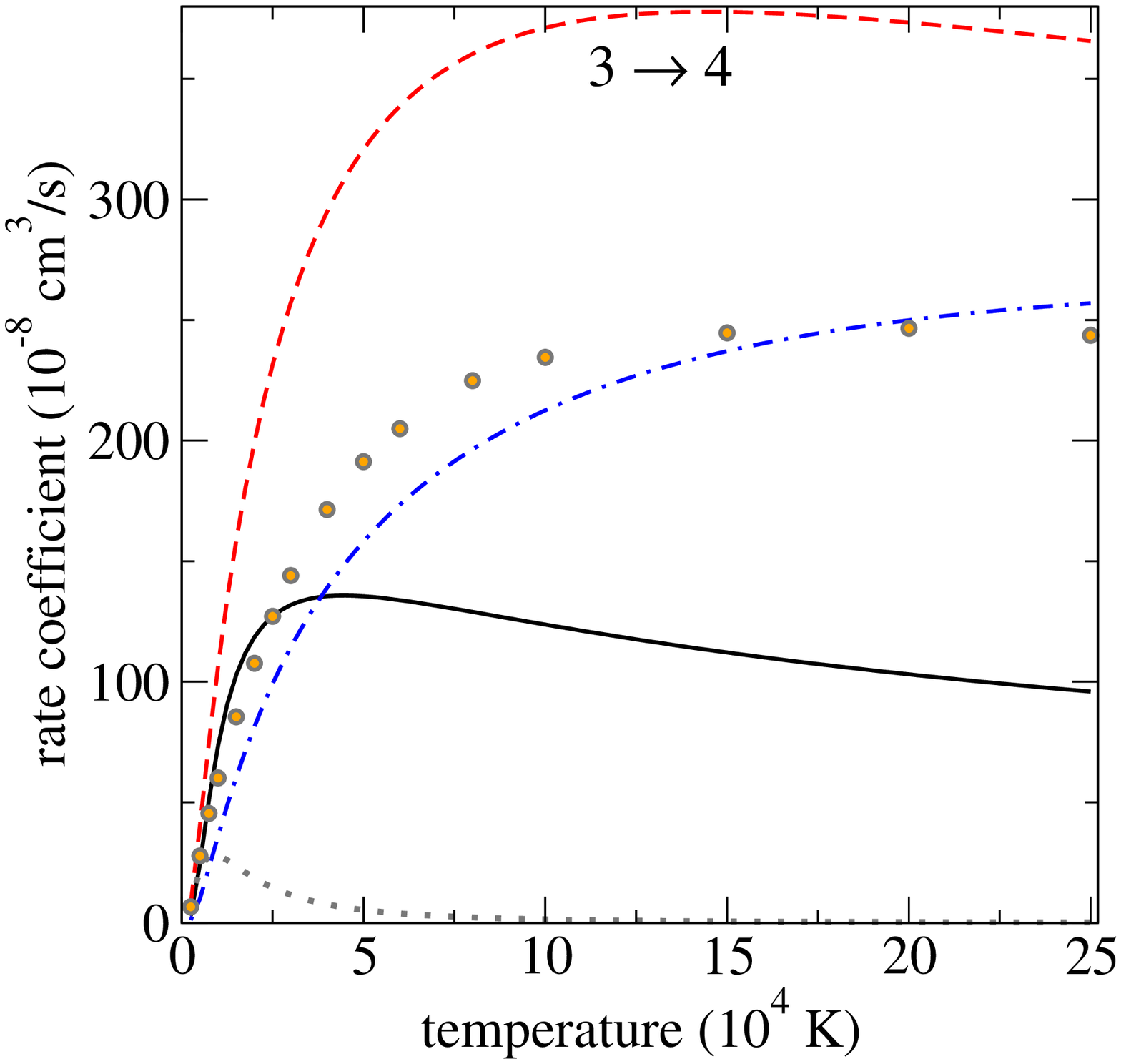}
\includegraphics[width=\figwidth]{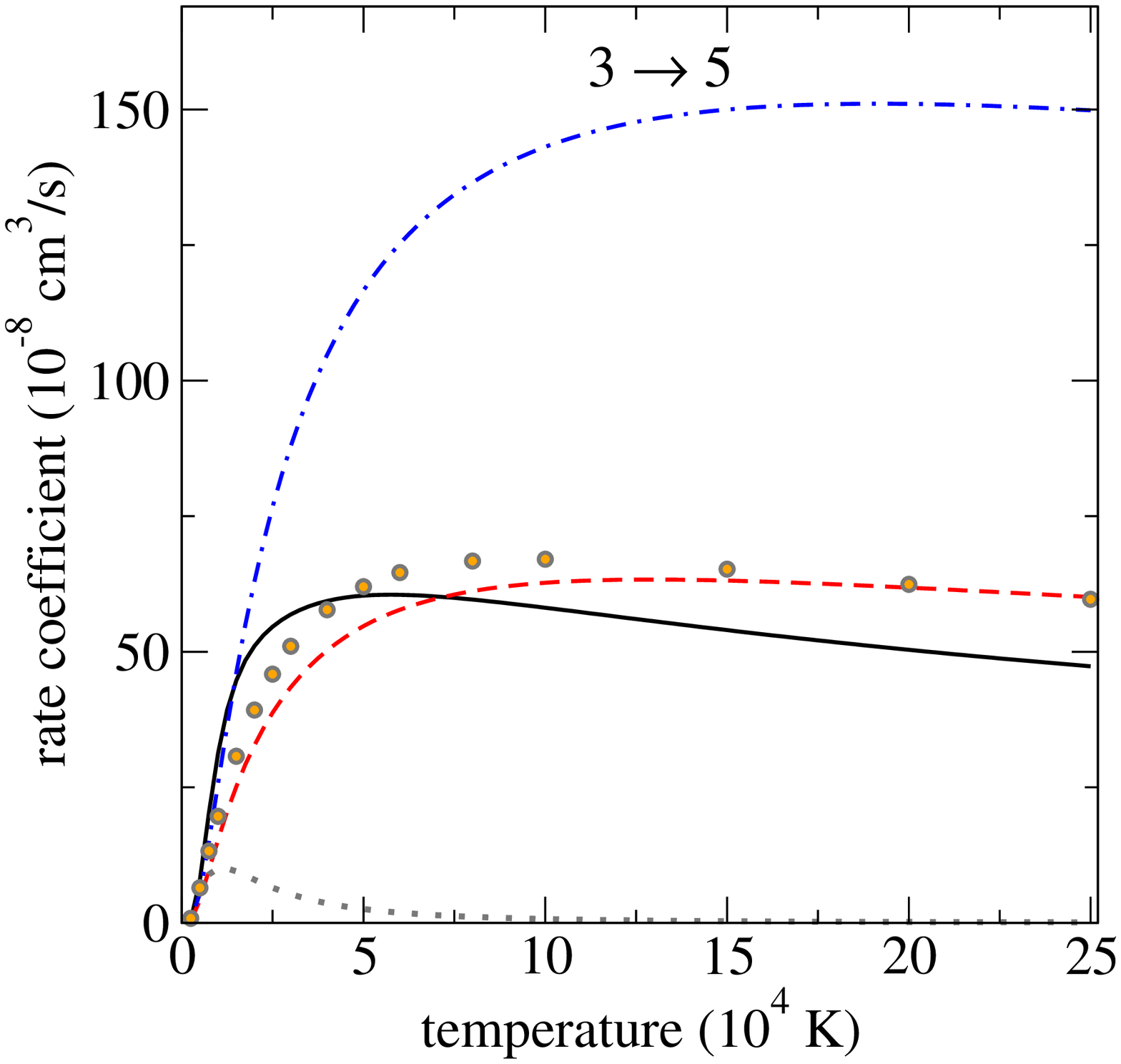}
\includegraphics[width=\figwidth]{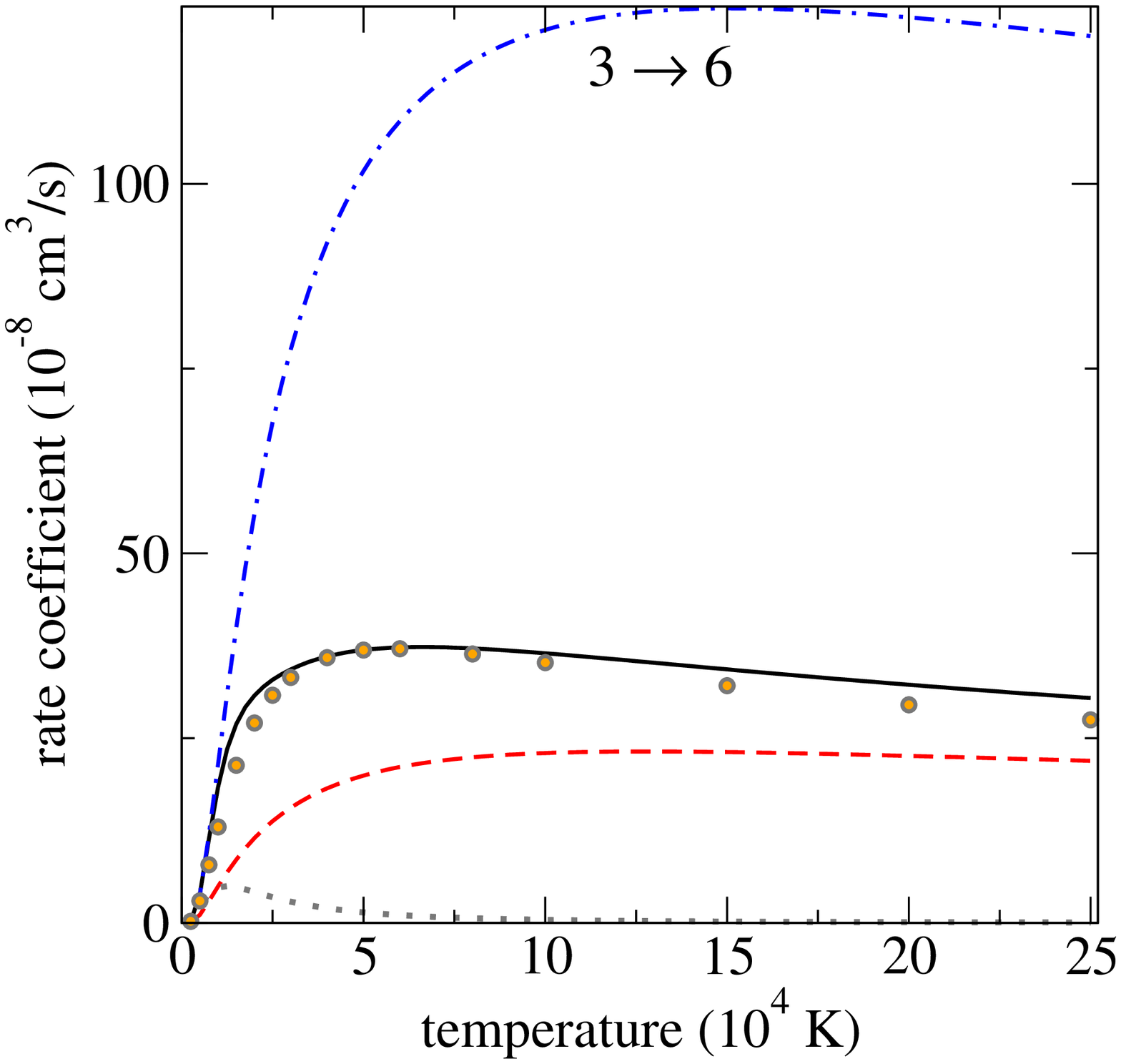}
\includegraphics[width=\figwidth]{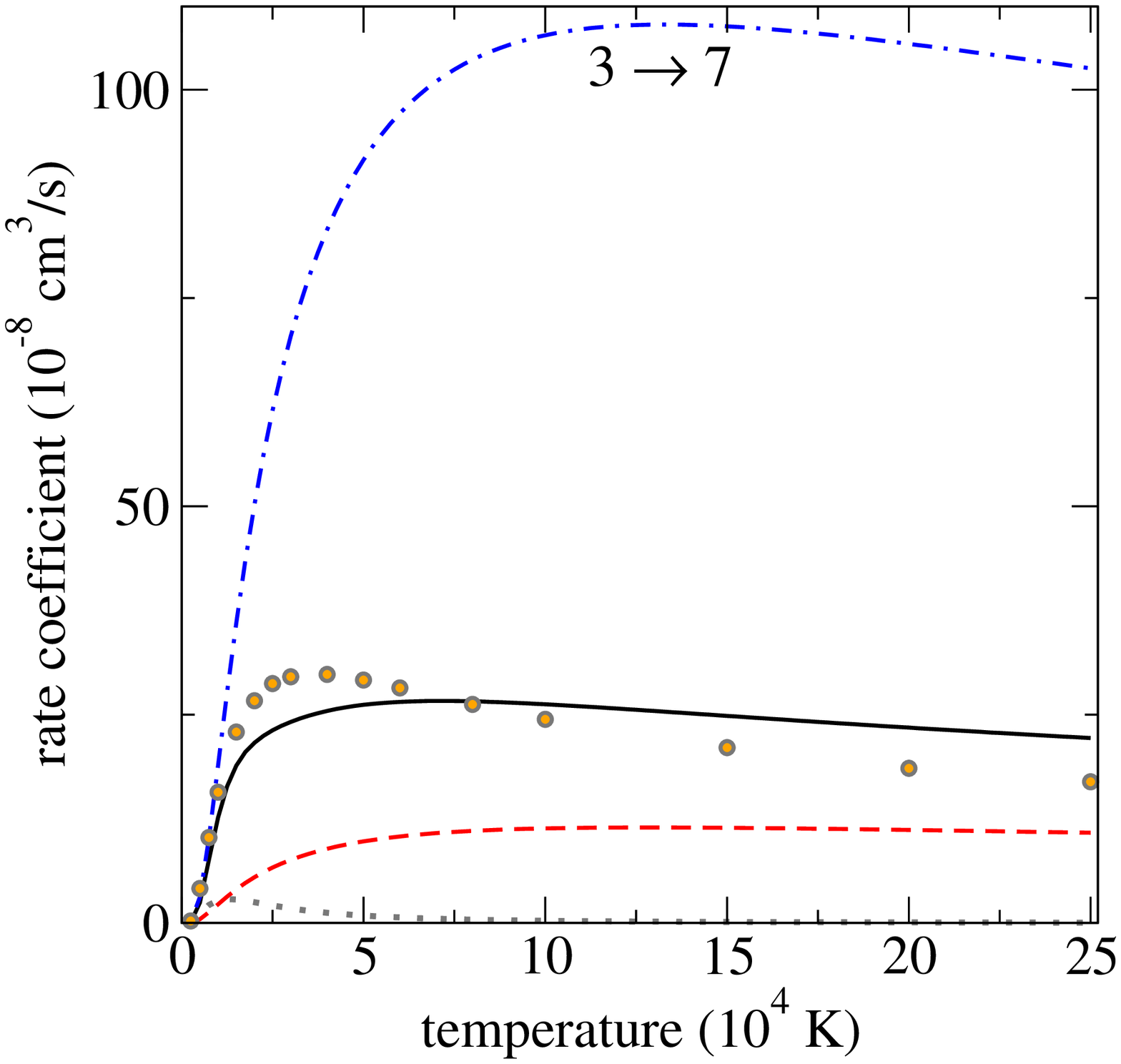}
\includegraphics[width=\figwidth]{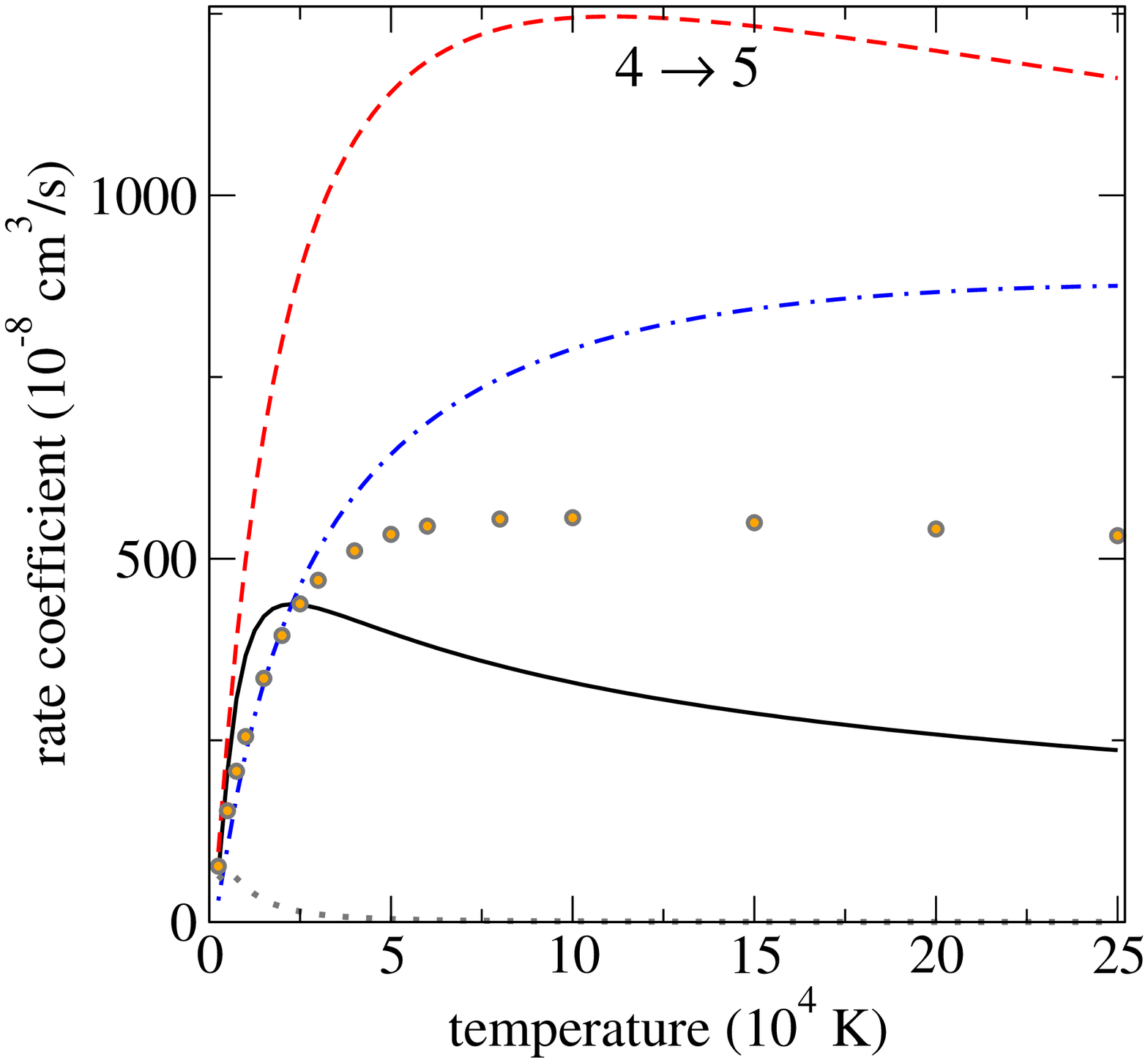}
\includegraphics[width=\figwidth]{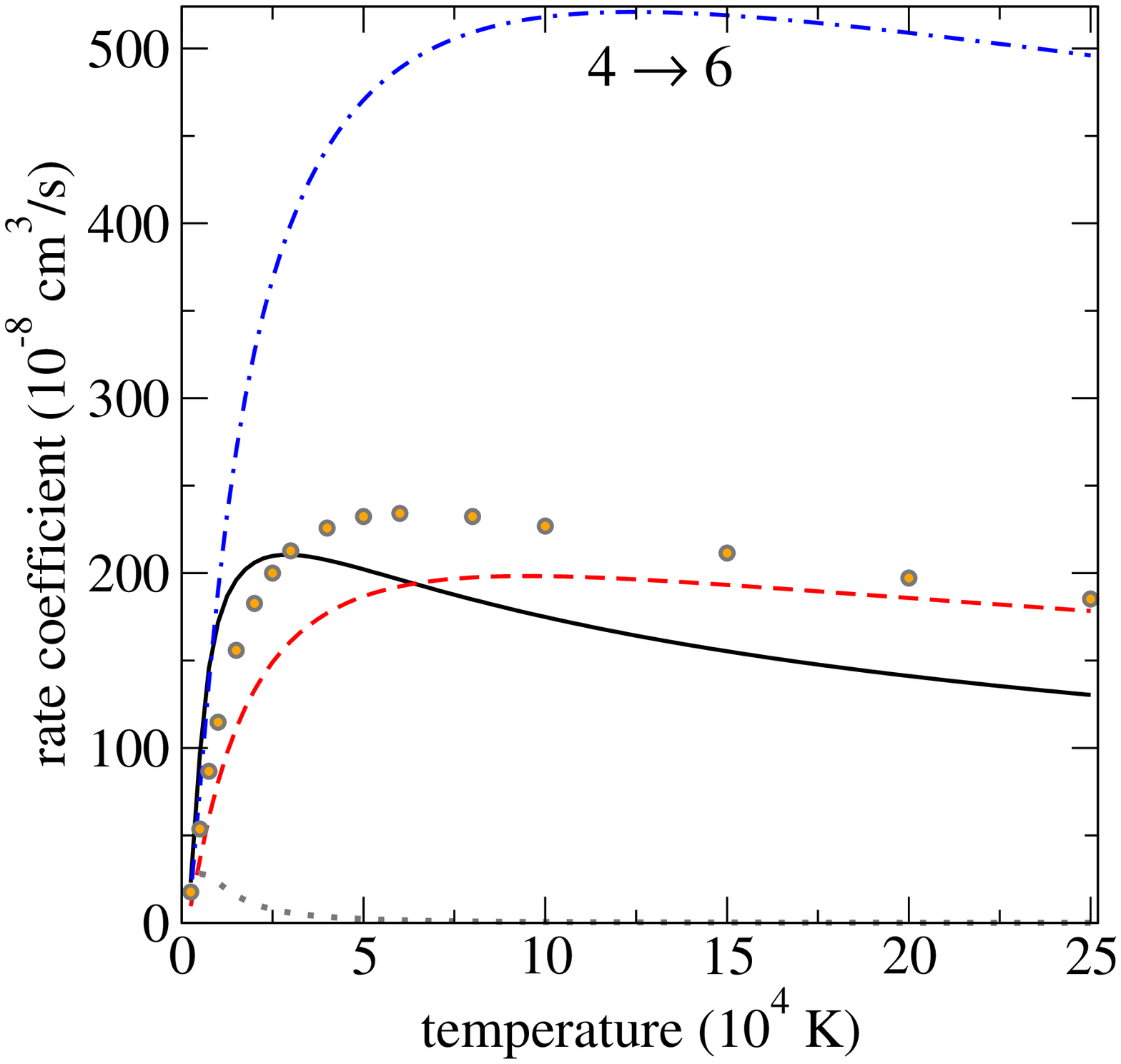}
\includegraphics[width=\figwidth]{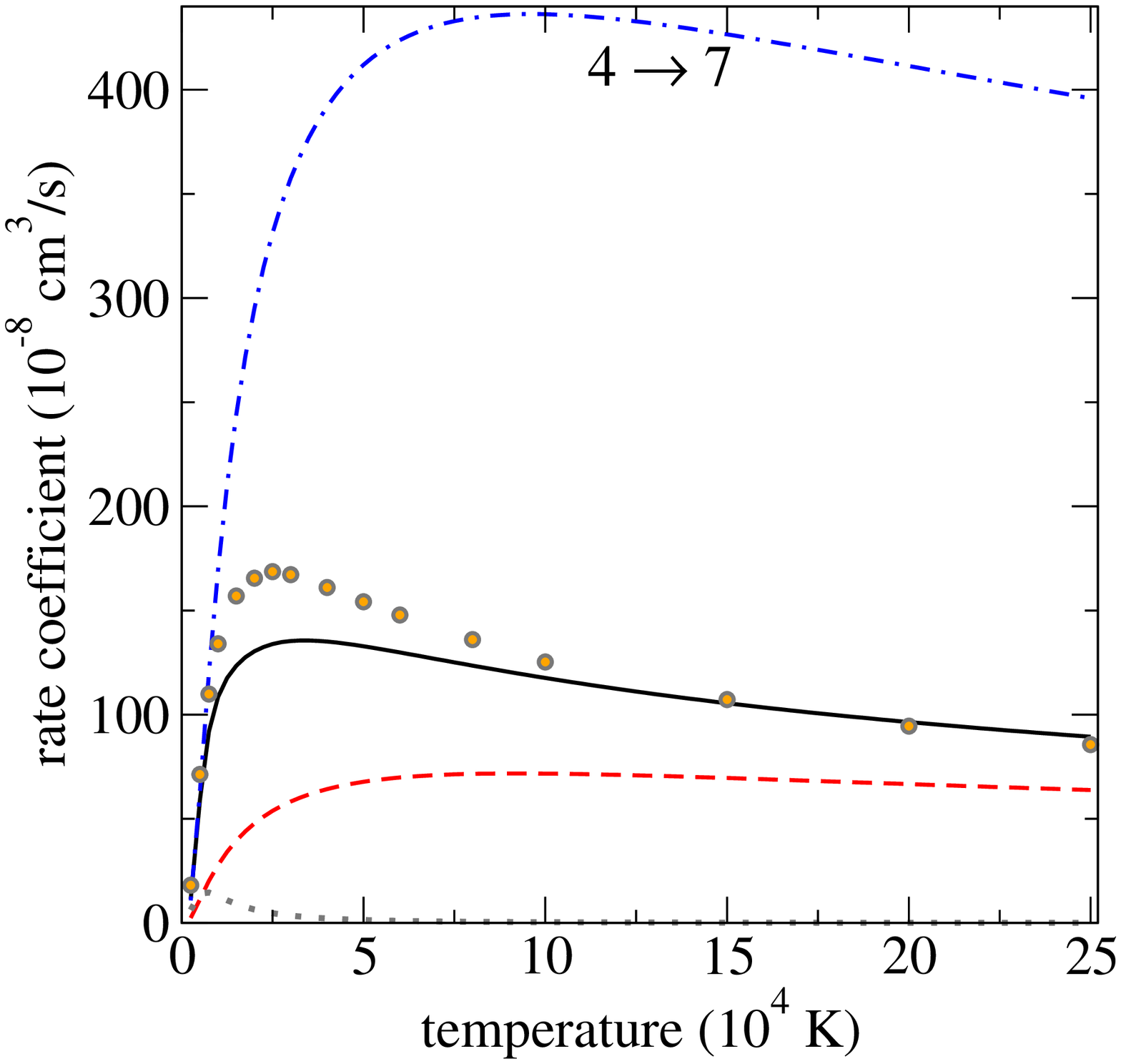}
\includegraphics[width=\figwidth]{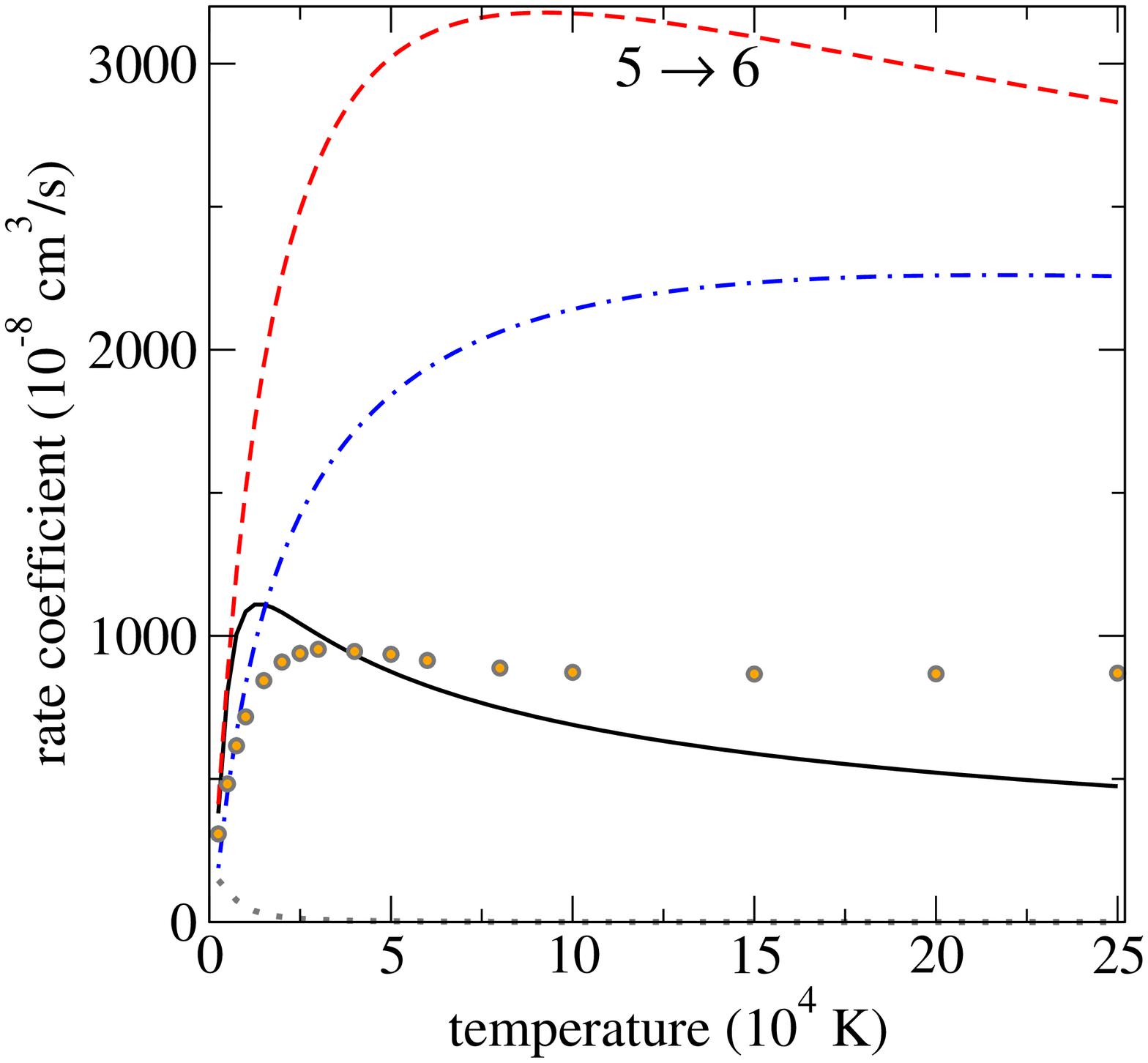}
\includegraphics[width=\figwidth]{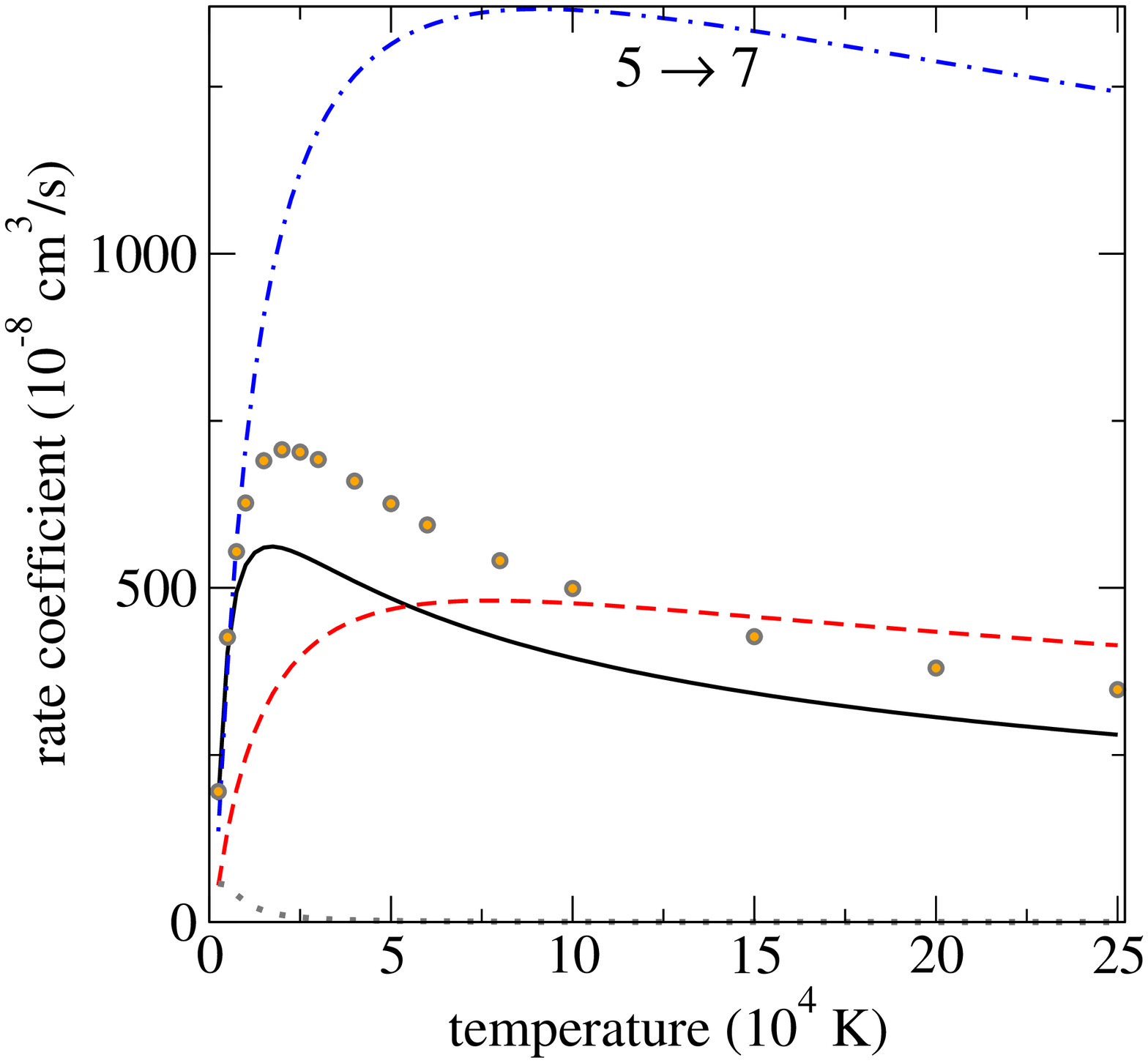}

\caption{Electron-hydrogen collision rate coefficients versus temperature 
for low level transitions $n_i \rightarrow n_f$ up to $n_i=5$ and $n_f=7$ as computed 
via Eq. (\ref{P2}) with solid lines, by using the semi-empirical formulas proposed 
by Johnson (\cite{johnson}) with dashed lines and by Beigman \& Lebedev (\cite{beigman})
with dash-dotted lines. Semiclassical results (Mansbach \& Keck \cite{mansbach}) are represented
by dotted lines and the R-matrix calculations of Przybilla \& Butler (\cite{przybilla}) are
marked by disks.}
\label{FigColl}
\end{figure*}

\begin{table}
\begin{center}
\caption{\label{table1}
Electron-hydrogen collision rate coefficients for specified transitions in 
$\mathrm{cm^3/s}$ at temperatures up to 30,000 K, calculated by the present formula 
(upper value) and by the R-matrix method (Pryzbilla \& Butler \cite{przybilla}) (lower value).}
{\footnotesize 
\begin{tabular}{|c|cccccccc|}
\tableline
\tableline
& \multicolumn{8}{c}{T[$10^4$ K]} \\
\cline{2-9}
Transition & 0.25 & 0.5 & 0.75 & 1 & 1.5 & 2 & 2.5 & 3 \\
\tableline
{$1\rightarrow 2$} & 3.148E-30 & 1.454E-19 & 6.373E-16 & 4.616E-14 & 3.739E-12 & 3.609E-11 & 1.455E-10 & 3.751E-10 \\
                   &1.484E-28 & 2.208E-18 & 5.244E-15 & 2.513E-13 & 1.178E-11 & 8.007E-11 & 2.537E-10 & 5.532E-10 \\
\hline
{$1\rightarrow 3$} & 1.753E-34 & 6.520E-22 & 1.236E-17 & 1.862E-15 & 3.143E-13 & 4.383E-12 & 2.204E-11 & 6.583E-11 \\
              &7.912E-33 & 9.454E-21 & 9.304E-17 & 9.049E-15 & 8.769E-13 & 8.806E-12 & 3.573E-11 & 9.171E-11 \\
\hline
{$1\rightarrow 4$} & 4.266E-36 & 7.368E-23 & 2.331E-18 & 4.540E-16 & 9.905E-14 & 1.571E-12 & 8.533E-12 & 2.684E-11 \\
              &1.658E-34 & 8.657E-22 & 1.471E-17 & 1.963E-15 & 2.756E-13 & 3.354E-12 & 1.508E-11 & 4.089E-11 \\
\hline
{$1\rightarrow 5$} & 6.595E-37 & 2.319E-23 & 9.298E-19 & 2.039E-16 & 5.011E-14 & 8.433E-13 & 4.748E-12 & 1.529E-11 \\
              &1.983E-35 & 2.436E-22 & 5.972E-18 & 9.679E-16 & 1.613E-13 & 2.075E-12 & 9.529E-12 & 2.612E-11 \\
\hline
{$1\rightarrow 6$} & 2.195E-37 & 1.136E-23 & 5.179E-19 & 1.212E-16 & 3.176E-14 & 5.520E-13 & 3.169E-12 & 1.034E-11 \\
              &5.533E-36 & 1.321E-22 & 4.044E-18 & 7.143E-16 & 1.244E-13 & 1.615E-12 & 7.351E-12 & 1.987E-11 \\
\hline
{$1\rightarrow 7$} & 1.070E-37 & 6.984E-24 & 3.442E-19 & 8.371E-17 & 2.281E-14 & 4.044E-13 & 2.349E-12 & 7.723E-12 \\
              &5.884E-36 & 1.510E-22 & 4.334E-18 & 7.236E-16 & 1.172E-13 & 1.470E-12 & 6.511E-12 & 1.739E-11 \\
\hline
{$2\rightarrow 3$} & 1.116E-11 & 1.861E-9 & 1.155E-8 & 2.986E-8 & 7.838E-8 & 1.250E-7 & 1.617E-7 & 1.887E-7 \\
              &7.864E-11 & 5.282E-9 & 2.068E-8 & 4.069E-8 & 8.187E-8 & 1.200E-7 & 1.547E-7 & 1.859E-7 \\
\hline
{$2\rightarrow 4$} & 2.769E-13 & 2.165E-10 & 2.251E-9 & 7.501E-9 & 2.504E-8 & 4.418E-8 & 5.972E-8 & 7.087E-8 \\
              &1.662E-12 & 4.707E-10 & 2.958E-9 & 7.488E-9 & 1.983E-8 & 3.299E-8 & 4.551E-8 & 5.665E-8 \\
\hline
{$2\rightarrow 5$} & 4.309E-14 & 6.881E-11 & 9.079E-10 & 3.406E-9 & 1.275E-8 & 2.369E-8 & 3.282E-8 & 3.941E-8 \\
              &1.958E-13 & 1.187E-10 & 1.042E-9 & 3.192E-9 & 1.004E-8 & 1.788E-8 & 2.518E-8 & 3.135E-8 \\
\hline
{$2\rightarrow 6$} & 1.438E-14 & 3.385E-11 & 5.085E-10 & 2.035E-9 & 8.112E-9 & 1.550E-8 & 2.180E-8 & 2.636E-8 \\
              &5.044E-14 & 6.195E-11 & 7.017E-10 & 2.364E-9 & 7.788E-9 & 1.386E-8 & 1.911E-8 & 2.340E-8 \\
\hline
{$2\rightarrow 7$} & 7.020E-15 & 2.087E-11 & 3.389E-10 & 1.410E-9 & 5.840E-9 & 1.136E-8 & 1.611E-8 & 1.957E-8 \\
              &5.714E-14 & 7.867E-11 & 8.399E-10 & 2.665E-9 & 8.094E-9 & 1.345E-8 & 1.792E-8 & 2.121E-8 \\
\hline
{$3\rightarrow 4$} & 2.996E-8 & 2.425E-7 & 5.092E-7 & 7.345E-7 & 1.029E-6 & 1.185E-6 & 1.271E-6 & 1.319E-6 \\
              &6.677E-8 & 2.776E-7 & 4.536E-7 & 6.009E-7 & 8.542E-7 & 1.076E-6 & 1.271E-6 & 1.440E-6 \\
\hline
{$3\rightarrow 5$} & 4.842E-9 & 7.905E-8 & 2.027E-7 & 3.130E-7 & 4.468E-7 & 5.106E-7 & 5.462E-7 & 5.688E-7 \\
              &8.481E-9 & 6.468E-8 & 1.326E-7 & 1.966E-7 & 3.074E-7 & 3.926E-7 & 4.586E-7 & 5.102E-7 \\
\hline
{$3\rightarrow 6$} & 1.638E-9 & 3.933E-8 & 1.135E-7 & 1.838E-7 & 2.691E-7 & 3.077E-7 & 3.289E-7 & 3.430E-7 \\
              &2.052E-9 & 2.982E-8 & 7.864E-8 & 1.299E-7 & 2.133E-7 & 2.704E-7 & 3.081E-7 & 3.321E-7 \\
\hline
{$3\rightarrow 7$} & 8.048E-10 & 2.440E-8 & 7.575E-8 & 1.265E-7 & 1.888E-7 & 2.163E-7 & 2.312E-7 & 2.412E-7 \\
              &2.370E-9 & 4.138E-8 & 1.025E-7 & 1.569E-7 & 2.290E-7 & 2.667E-7 & 2.873E-7 & 2.954E-7 \\
\hline
{$4\rightarrow 5$} & 6.634E-7 & 2.074E-6 & 3.070E-6 & 3.669E-6 & 4.206E-6 & 4.359E-6 & 4.367E-6 & 4.316E-6 \\
              &7.685E-7 & 1.531E-6 & 2.075E-6 & 2.552E-6 & 3.354E-6 & 3.944E-6 & 4.380E-6 & 4.703E-6 \\
\hline
{$4\rightarrow 6$} & 2.307E-7 & 9.656E-7 & 1.458E-6 & 1.720E-6 & 1.961E-6 & 2.060E-6 & 2.098E-6 & 2.105E-6 \\
              &1.770E-7 & 5.364E-7 & 8.666E-7 & 1.147E-6 & 1.558E-6 & 1.827E-6 & 2.000E-6 & 2.127E-6 \\
\hline
{$4\rightarrow 7$} & 1.148E-7 & 5.889E-7 & 9.191E-7 & 1.085E-6 & 1.235E-6 & 1.304E-6 & 1.339E-6 & 1.354E-6 \\
              &1.810E-7 & 7.128E-7 & 1.099E-6 & 1.340E-6 & 1.569E-6 & 1.655E-6 & 1.687E-6 & 1.672E-6 \\
\hline
{$5\rightarrow 6$} & 3.801E-6 & 8.054E-6 & 1.006E-5 & 1.085E-5 & 1.110E-5 & 1.082E-5 & 1.043E-5 & 1.004E-5 \\
              &3.079E-6 & 4.829E-6 & 6.164E-6 & 7.172E-6 & 8.439E-6 & 9.087E-6 & 9.388E-6 & 9.531E-6 \\
\hline
{$5\rightarrow 7$} & 1.822E-6 & 4.033E-6 & 4.939E-6 & 5.341E-6 & 5.604E-6 & 5.597E-6 & 5.499E-6 & 5.370E-6 \\
              &1.953E-6 & 4.260E-6 & 5.542E-6 & 6.271E-6 & 6.903E-6 & 7.068E-6 & 7.032E-6 & 6.921E-6 \\
\hline
{$6\rightarrow 7$} & 1.298E-5 & 2.175E-5 & 2.401E-5 & 2.419E-5 & 2.309E-5 & 2.175E-5 & 2.052E-5 & 1.947E-5 \\
              &1.025E-5 & 1.571E-5 & 1.778E-5 & 1.846E-5 & 1.838E-5 & 1.767E-5 & 1.686E-5 & 1.604E-5 \\
\tableline
\end{tabular}
}

\end{center}
\end{table}

\begin{table}
\begin{center}
\caption{\label{table2}
Electron-hydrogen collision rate coefficients for specified transitions in 
$\mathrm{cm^3/s}$ at temperatures greater than 30,000 K, calculated by the present formula 
(upper value) and by the R-matrix method (Pryzbilla \& Butler \cite{przybilla}) (lower value).}

{\footnotesize

\begin{tabular}{|c|cccccccc|}
\tableline
\tableline
& \multicolumn{8}{c}{T[$10^4$ K]} \\
\cline{2-9}
Transition& 4 & 5 & 6 & 8 & 10 & 15 & 20 & 25 \\
\tableline
\tableline
{$1\rightarrow 2$} & 1.250E-9 & 2.587E-9 & 4.173E-9 & 7.363E-9 & 9.954E-9 & 1.358E-8 & 1.516E-8 & 1.607E-8 \\
              &1.475E-9 & 2.729E-9 & 4.112E-9 & 7.014E-9 & 9.729E-9 & 1.503E-8 & 1.868E-8 & 2.123E-8 \\
\hline
{$1\rightarrow 3$} & 2.634E-10 & 6.071E-10 & 1.050E-9 & 2.007E-9 & 2.822E-9 & 3.966E-9 & 4.438E-9 & 4.706E-9 \\
              &2.997E-10 & 6.130E-10 & 9.834E-10 & 1.769E-9 & 2.492E-9 & 3.847E-9 & 4.683E-9 & 5.218E-9 \\
\hline
{$1\rightarrow 4$} & 1.145E-10 & 2.741E-10 & 4.860E-10 & 9.561E-10 & 1.364E-9 & 1.942E-9 & 2.176E-9 & 2.309E-9 \\
              &1.418E-10 & 2.949E-10 & 4.753E-10 & 8.514E-10 & 1.189E-9 & 1.786E-9 & 2.131E-9 & 2.330E-9 \\
\hline
{$1\rightarrow 5$} & 6.722E-11 & 1.638E-10 & 2.937E-10 & 5.857E-10 & 8.417E-10 & 1.205E-9 & 1.352E-9 & 1.435E-9 \\
              &9.025E-11 & 1.871E-10 & 2.986E-10 & 5.253E-10 & 7.163E-10 & 1.050E-9 & 1.225E-9 & 1.323E-9 \\
\hline
{$1\rightarrow 6$} & 4.619E-11 & 1.136E-10 & 2.051E-10 & 4.119E-10 & 5.942E-10 & 8.538E-10 & 9.583E-10 & 1.017E-9 \\
              &6.741E-11 & 1.371E-10 & 2.155E-10 & 3.695E-10 & 4.998E-10 & 7.087E-10 & 8.151E-10 & 8.640E-10 \\
\hline
{$1\rightarrow 7$} & 3.484E-11 & 8.620E-11 & 1.562E-10 & 3.151E-10 & 4.556E-10 & 6.559E-10 & 7.365E-10 & 7.818E-10 \\
              &5.735E-11 & 1.146E-10 & 1.793E-10 & 3.002E-10 & 3.981E-10 & 5.523E-10 & 6.189E-10 & 6.508E-10 \\
\hline
{$2\rightarrow 3$} & 2.226E-7 & 2.419E-7 & 2.539E-7 & 2.667E-7 & 2.717E-7 & 2.700E-7 & 2.617E-7 & 2.522E-7 \\
              &2.404E-7 & 2.844E-7 & 3.209E-7 & 3.740E-7 & 4.137E-7 & 4.645E-7 & 4.885E-7 & 4.980E-7 \\
\hline
{$2\rightarrow 4$} & 8.397E-8 & 9.097E-8 & 9.549E-8 & 1.012E-7 & 1.044E-7 & 1.066E-7 & 1.054E-7 & 1.029E-7 \\
              &7.437E-8 & 8.727E-8 & 9.680E-8 & 1.091E-7 & 1.157E-7 & 1.214E-7 & 1.213E-7 & 1.194E-7 \\
\hline
{$2\rightarrow 5$} & 4.702E-8 & 5.097E-8 & 5.351E-8 & 5.688E-8 & 5.896E-8 & 6.096E-8 & 6.073E-8 & 5.966E-8 \\
              &4.097E-8 & 4.723E-8 & 5.145E-8 & 5.619E-8 & 5.803E-8 & 5.828E-8 & 5.681E-8 & 5.461E-8 \\
\hline
{$2\rightarrow 6$} & 3.160E-8 & 3.428E-8 & 3.599E-8 & 3.832E-8 & 3.983E-8 & 4.143E-8 & 4.145E-8 & 4.085E-8 \\
              &2.939E-8 & 3.300E-8 & 3.534E-8 & 3.714E-8 & 3.747E-8 & 3.593E-8 & 3.400E-8 & 3.206E-8 \\
\hline
{$2\rightarrow 7$} & 2.354E-8 & 2.555E-8 & 2.684E-8 & 2.860E-8 & 2.977E-8 & 3.108E-8 & 3.117E-8 & 3.077E-8 \\
              &2.550E-8 & 2.757E-8 & 2.864E-8 & 2.885E-8 & 2.849E-8 & 2.603E-8 & 2.395E-8 & 2.221E-8 \\
\hline
{$3\rightarrow 4$} & 1.355E-6 & 1.355E-6 & 1.338E-6 & 1.290E-6 & 1.237E-6 & 1.121E-6 & 1.031E-6 & 9.590E-7 \\
              &1.714E-6 & 1.913E-6 & 2.050E-6 & 2.248E-6 & 2.345E-6 & 2.446E-6 & 2.466E-6 & 2.436E-6 \\
\hline
{$3\rightarrow 5$} & 5.938E-7 & 6.035E-7 & 6.051E-7 & 5.963E-7 & 5.812E-7 & 5.400E-7 & 5.036E-7 & 4.732E-7 \\
              &5.776E-7 & 6.201E-7 & 6.461E-7 & 6.673E-7 & 6.708E-7 & 6.524E-7 & 6.243E-7 & 5.968E-7 \\
\hline
{$3\rightarrow 6$} & 3.605E-7 & 3.693E-7 & 3.728E-7 & 3.714E-7 & 3.648E-7 & 3.431E-7 & 3.223E-7 & 3.043E-7 \\
              &3.589E-7 & 3.692E-7 & 3.710E-7 & 3.638E-7 & 3.522E-7 & 3.209E-7 & 2.951E-7 & 2.747E-7 \\
\hline
{$3\rightarrow 7$} & 2.544E-7 & 2.618E-7 & 2.653E-7 & 2.659E-7 & 2.624E-7 & 2.485E-7 & 2.344E-7 & 2.219E-7 \\
              &2.983E-7 & 2.914E-7 & 2.821E-7 & 2.622E-7 & 2.444E-7 & 2.104E-7 & 1.856E-7 & 1.693E-7 \\
\hline
{$4\rightarrow 5$} & 4.153E-6 & 3.978E-6 & 3.813E-6 & 3.527E-6 & 3.295E-6 & 2.872E-6 & 2.581E-6 & 2.365E-6 \\
              &5.108E-6 & 5.336E-6 & 5.448E-6 & 5.545E-6 & 5.564E-6 & 5.495E-6 & 5.409E-6 & 5.318E-6 \\
\hline
{$4\rightarrow 6$} & 2.075E-6 & 2.022E-6 & 1.963E-6 & 1.849E-6 & 1.749E-6 & 1.553E-6 & 1.412E-6 & 1.304E-6 \\
              &2.257E-6 & 2.324E-6 & 2.341E-6 & 2.324E-6 & 2.269E-6 & 2.115E-6 & 1.971E-6 & 1.852E-6 \\
\hline
{$4\rightarrow 7$} & 1.351E-6 & 1.328E-6 & 1.299E-6 & 1.235E-6 & 1.176E-6 & 1.055E-6 & 9.643E-7 & 8.940E-7 \\
              &1.610E-6 & 1.542E-6 & 1.478E-6 & 1.360E-6 & 1.253E-6 & 1.073E-6 & 9.450E-7 & 8.561E-7 \\
\hline
{$5\rightarrow 6$} & 9.339E-6 & 8.750E-6 & 8.258E-6 & 7.483E-6 & 6.895E-6 & 5.883E-6 & 5.221E-6 & 4.744E-6 \\
              &9.457E-6 & 9.358E-6 & 9.143E-6 & 8.876E-6 & 8.727E-6 & 8.667E-6 & 8.677E-6 & 8.701E-6 \\
\hline
{$5\rightarrow 7$} & 5.097E-6 & 4.843E-6 & 4.618E-6 & 4.245E-6 & 3.950E-6 & 3.422E-6 & 3.066E-6 & 2.803E-6 \\
              &6.598E-6 & 6.262E-6 & 5.943E-6 & 5.407E-6 & 4.990E-6 & 4.270E-6 & 3.800E-6 & 3.478E-6 \\ 
\hline
{$6\rightarrow 7$} & 1.775E-5 & 1.642E-5 & 1.536E-5 & 1.374E-5 & 1.255E-5 & 1.057E-5 & 9.308E-6 & 8.411E-6 \\
              &1.467E-5 & 1.362E-5 & 1.291E-5 & 1.211E-5 & 1.188E-5 & 1.210E-5 & 1.260E-5 & 1.312E-5 \\
\tableline

\end{tabular}
}
\end{center}
\end{table}

\end{document}